\documentclass[a4paper,fleqn,usenatbib]{mnras}
\usepackage{rotating}
\usepackage{lscape}
\usepackage{morefloats}
\usepackage{amssymb}
\DeclareMathAlphabet{\mathpzc}{OT1}{pzc}{m}{it}
\usepackage{amsmath}
\usepackage{longtable}

\def\lsim{\,\lower2truept\hbox{${<\atop\hbox{\raise4truept\hbox{$\sim$}}}$}\,}
\def\gsim{\,\lower2truept\hbox{${> \atop\hbox{\raise4truept\hbox{$\sim$}}}$}\,}

\title[1.4 GHz luminosity functions]{New constraints on the 1.4 GHz source number
counts and luminosity functions in the Lockman Hole field}
\author[M. Bonato et al.]
{Matteo Bonato$^{1,2}$\thanks{matteo.bonato@inaf.it}, Isabella Prandoni$^{3}$, Gianfranco De Zotti$^{2}$, Marisa Brienza$^{4,3}$, \newauthor Raffaella Morganti$^{5,6}$
 and Mattia Vaccari$^{7,8,3}$\\
$^{1}$INAF$-$Istituto di Radioastronomia and Italian ALMA Regional Centre, Via Gobetti 101, I-40129, Bologna, Italy\\
$^{2}$INAF$-$Osservatorio Astronomico di Padova, Vicolo dell'Osservatorio 5, I-35122, Padova, Italy\\
$^{3}$INAF$-$Istituto di Radioastronomia, Via Gobetti 101, I-40129, Bologna, Italy\\
$^{4}$Dipartimento di Fisica e Astronomia, Universit\'a di Bologna, via P. Gobetti 93/2, 40129, Bologna, Italy\\
$^{5}$Kapteyn Astronomical Institute, University of Groningen, PO Box 800, 9700 AV, Groningen, The Netherlands\\
$^{6}$ASTRON, Netherlands Institute for Radio Astronomy, Oude Hoogeveensedijk 4, 7991 PD, Dwingeloo, The Netherlands\\
$^{7}$Department of Physics and Astronomy, University of the Western Cape, Private Bag X17, Bellville 7535, South Africa\\
$^{8}$Inter-University Institute for Data Intensive Astronomy, South Africa}
\date{Released 2020 Xxxxx XX}

\pagerange{\pageref{firstpage}--\pageref{lastpage}} \pubyear{2020}

\def\LaTeX{L\kern-.36em\raise.3ex\hbox{a}\kern-.15em
    T\kern-.1667em\lower.7ex\hbox{E}\kern-.125emX}
\def\simlt{\mathrel{\rlap{\lower 3pt\hbox{$\sim$}}\raise 2.0pt\hbox{$<$}}}
\def\simgt{\mathrel{\rlap{\lower 3pt\hbox{$\sim$}}\raise 2.0pt\hbox{$>$}}}

\begin{document}

\label{firstpage}

\maketitle

\begin{abstract}
We present a study of the 1173 sources brighter than $S_{1.4\,\rm GHz}=
120\,\mu$Jy detected over an area of $\simeq 1.4\,\hbox{deg}^{2}$ in the
Lockman Hole field. Exploiting the multi-band information available in this
field for $\sim$79\% of the sample, sources have been classified into radio
loud (RL) active galactic nuclei (AGNs), star forming galaxies (SFGs) and
radio quiet (RQ) AGNs, using a variety of diagnostics available in the
literature. Exploiting the observed tight anti-correlations between IRAC
band\,1 or band\,2 and the source redshift we could assign a redshift to 177
sources missing a spectroscopic measurement or a reliable photometric
estimate. A Monte Carlo approach was used to take into account the spread
around the mean relation. The derived differential number counts and
luminosity functions at several redshifts of each population show a good
consistency with models and with earlier estimates made using data from
different surveys and applying different approaches. Our results confirm that
below $\sim300\,\mu$Jy SFGs$+$RQ AGNs overtake RL AGNs that dominate at
brighter flux densities.
We also confirm earlier indications of a similar evolution of RQ AGNs and SFGs. Finally, we
discuss the angular correlation function of our sources and highlight its
sensitivity to the criteria used for the classification.
\end{abstract}


\section{Introduction}\label{sec:introduction}

The advent of sub-mJy radio surveys has opened up a new view of the
extragalactic radio sky (see \citealt{DeZotti2010} and \citealt{Padovani2016}
for reviews). At brighter flux densities the radio emission is almost
exclusively of nuclear origin; such sources will be referred to as radio-loud
active galactic nuclei (RL AGNs). Fainter sources comprise an increasing
fraction of star-forming galaxies (SFGs) whose low-frequency radio emission is
dominated by synchrotron radiation from relativistic electrons, accelerated by
supernovae and their remnants, interacting with the galactic magnetic field. A
substantial fraction of sources detected by deep radio surveys also contain
active nuclei with weak radio emission, referred to as radio-quiet (RQ) AGNs.
The origin of the radio emission of these objects is still debated since in
most cases it is difficult to disentangle it from that of the host galaxies
(see discussions in \citealt{Bonato2017}, \citealt{Mancuso2017},
\citealt{Prandoni2018}, \citealt{Ceraj2018}).

Studies of the composition of the faint radio source populations have
been carried out exploiting deep radio surveys of fields for which a wealth of
multifrequency data is available, such as the Extended Chandra Deep Field-South
\citep[ECDFS;][]{Bonzini2013}, the Cosmic Evolution Survey (COSMOS) field
\citep{Smolcic2017b} and the European Large-Area ISO Survey-North\,1
(ELAIS-N1; \citealt{Ocran2017, Ocran20}), among others. Despite the rapidly growing
knowledge on this topic, considerable uncertainties remain and no general
consensus has been reached. Analyses of new samples are necessary to improve
the statistics and to pinpoint the effect of different classification criteria.

The Lockman Hole (LH; \citealt{Lockman1986}, \citealt{Lonsdale2003}) field is
one of the best-studied extra-galactic regions of the sky (see,
e.g., \citealt{Prandoni2018} for an overview of the available multi-band information).
It is fully covered by the SERVS Data Fusion\footnote{\url{http://www.mattiavaccari.net/df/}} (\citealt{Vaccari2015}, \citealt{Mauduit2012}), including flux density
measurements in the \textit{Spitzer} IRAC band 1 (down to the detection limit
of 2.21 $\mu$Jy), band 2 (down to 2.7 $\mu$Jy), band 3 (down to
40.8 $\mu$Jy), band 4 (down to 44.4 $\mu$Jy), in MIPS 24 $\mu$m band
(down to 286.6 $\mu$Jy) and in $K_{s}$ band (up to Vega
magnitude 21.5). The LH region was also the target of deep X-ray observations with the ROSAT,
the XMM-\textit{Newton} and the Chandra satellites (see Fig.\,1 of \citealt{Prandoni2018} and text for details).


\citet{Prandoni2018} presented  new Westerbork Synthesis Radio Telescope (WSRT)
1.4\,GHz mosaic observations, covering an area of 6.6 square degrees down to an
rms noise of $11\,\mu$Jy/beam, and the derived source counts. A comparison with
existing radio source evolutionary models was also presented, based on a
preliminary classification of the sources in the central 1.4\,deg$^{2}$. In
this paper, again focusing on the central $\sim$1.4\,deg$^{2}$, we
exploit the new radio data together with the available multi-band observations
to get robust observational constraints on star formation and AGN activity of
radio-detected sources, on their cosmic evolution and on their clustering
properties, in preparation for the deeper and wider continuum extra-galactic
surveys that will be carried out by the Square Kilometre Array (SKA).

In Sect.~\ref{sec:data} we briefly introduce our sample, present a
revised version of the source catalogue based on visual inspection of radio
source/host galaxy identifications and describe the classification of the
detected sources as either RL AGNs or SFGs or RQ AGNs. In
Sect.~\ref{sec:results} we present new estimates of the number counts of each
sub-population, of luminosity functions in 13 redshift bins and of the 2-point
angular correlation function. Our conclusions are summarized in
Sect.~\ref{sec:conclusions}. Throughout this paper, we adopt a flat $\Lambda
\rm CDM$ cosmology with $\Omega_{\rm m} = 0.31$, $\Omega_{\Lambda} = 0.69$ and
$h=H_0/100\, \rm km\,s^{-1}\,Mpc^{-1} = 0.67$ \citep{Planck2015}.

\section{Data and classification}\label{sec:data}

\begin{figure}
\begin{center}
\includegraphics[width=0.49\textwidth]{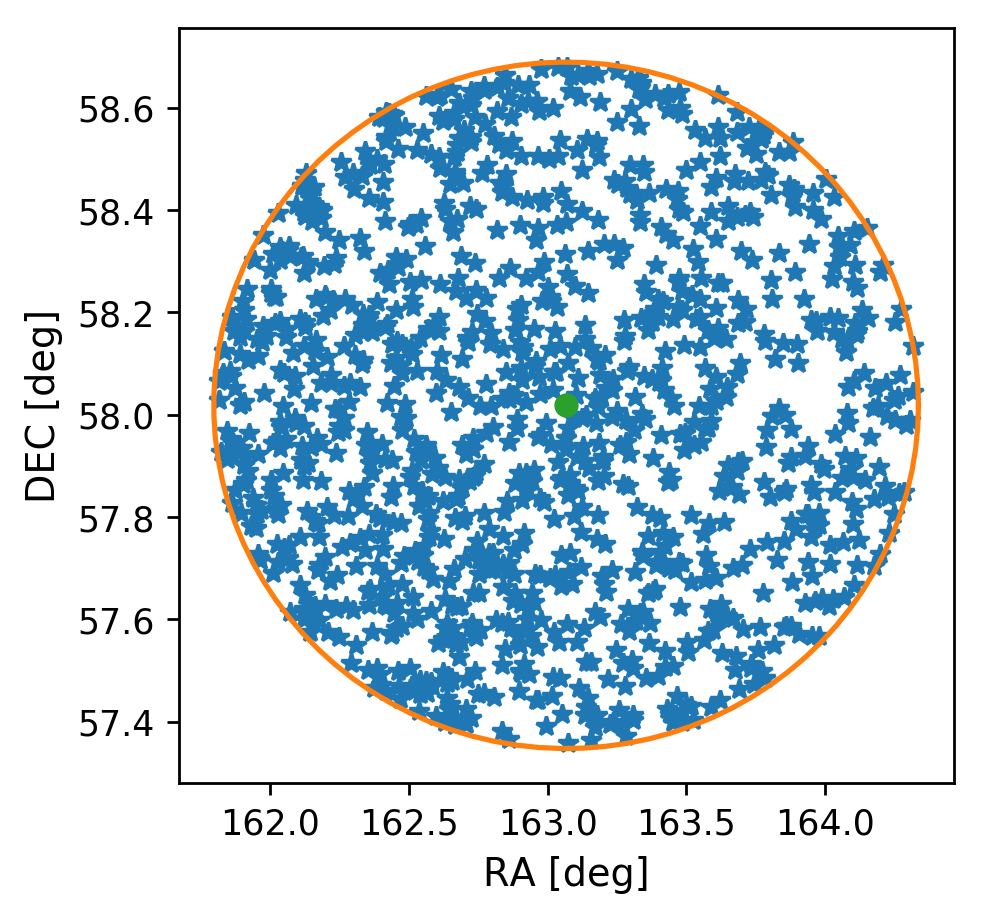}
\caption{Spatial distribution (in equatorial coordinates, J2000) of the sources
(the blue stars) analysed in this paper. The green central filled circle represents the
center of the region.}
 \label{fig:area}
  \end{center}
\end{figure}

\begin{figure*}
\begin{center}
\includegraphics[width=0.49\textwidth]{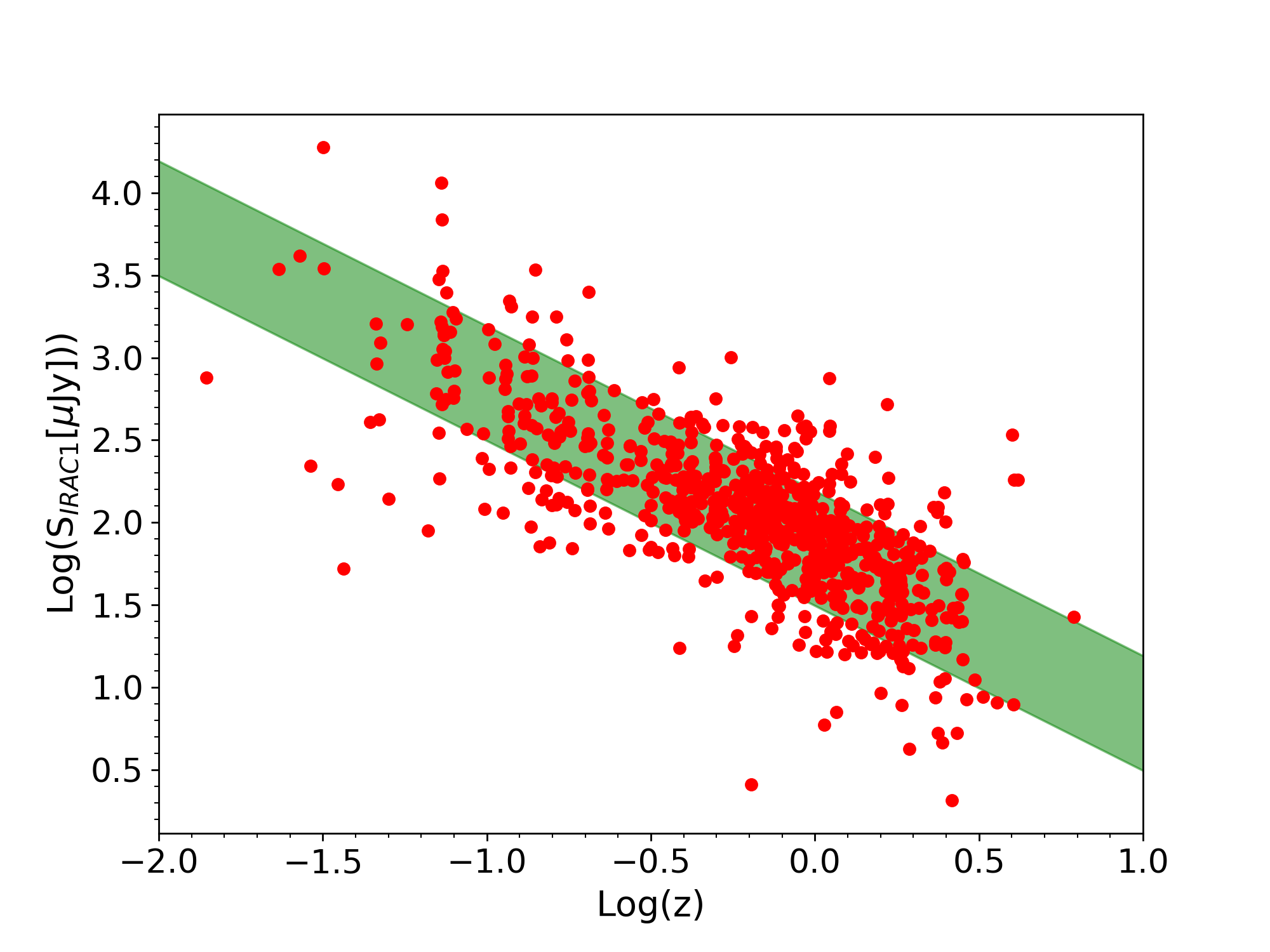}
\includegraphics[width=0.49\textwidth]{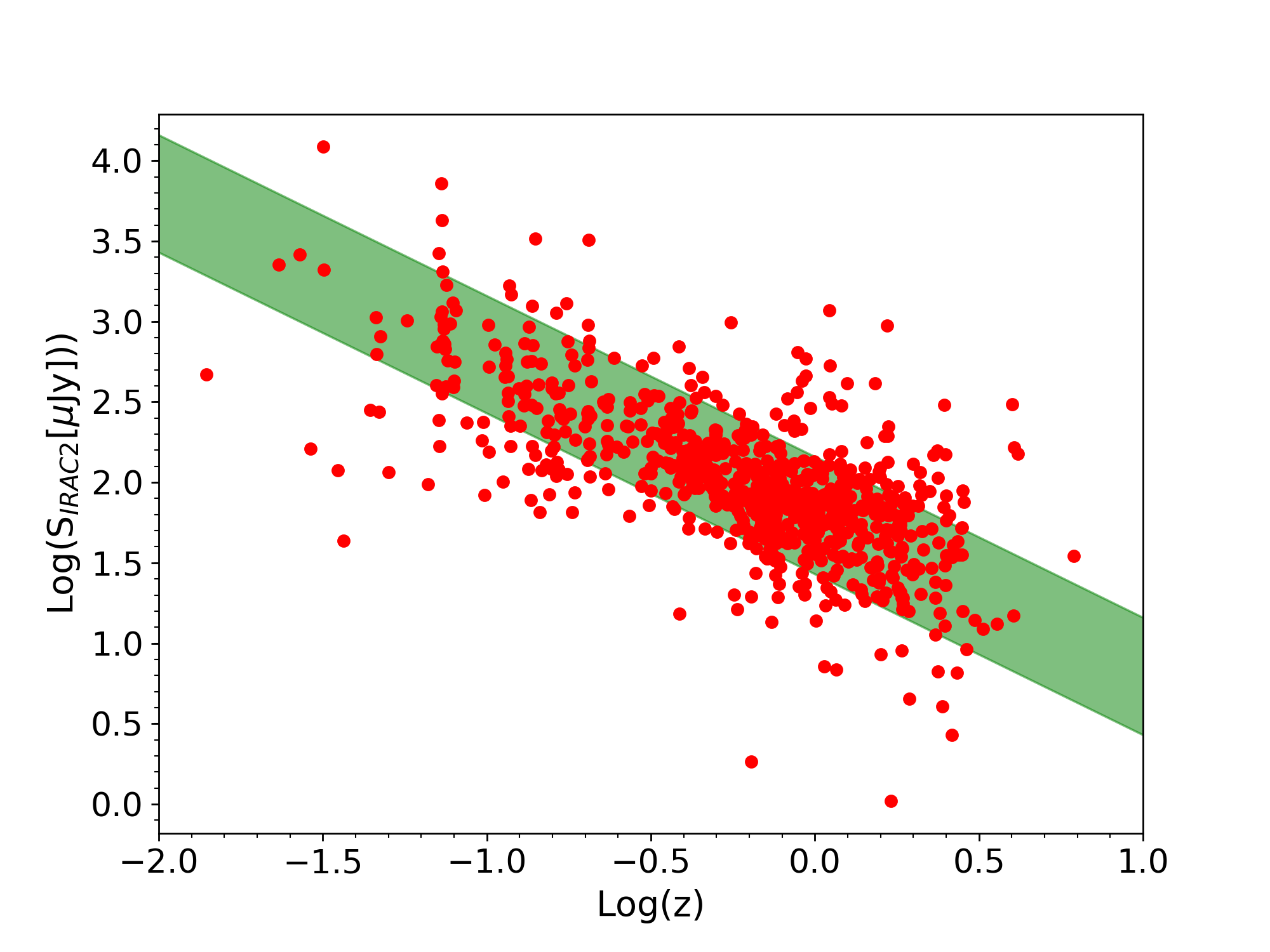}
\caption{IRAC band 1 (left) and band 2 (right) flux densities vs. redshift for sources with spectroscopic or
reliable photometric redshifts, in our sample. The green bands show the
$2\,\sigma$ range around the mean relation $\log S_{\rm IRAC}=-\log z+c_{\rm IRAC}$, with
$c_{\rm IRAC}= 1.79 \pm 0.36$ for band 2 and $=1.84 \pm 0.35$ for band 1.}
 \label{fig:correlations}
  \end{center}
\end{figure*}

\begin{figure*}
\begin{center}
\includegraphics[width=0.49\textwidth]{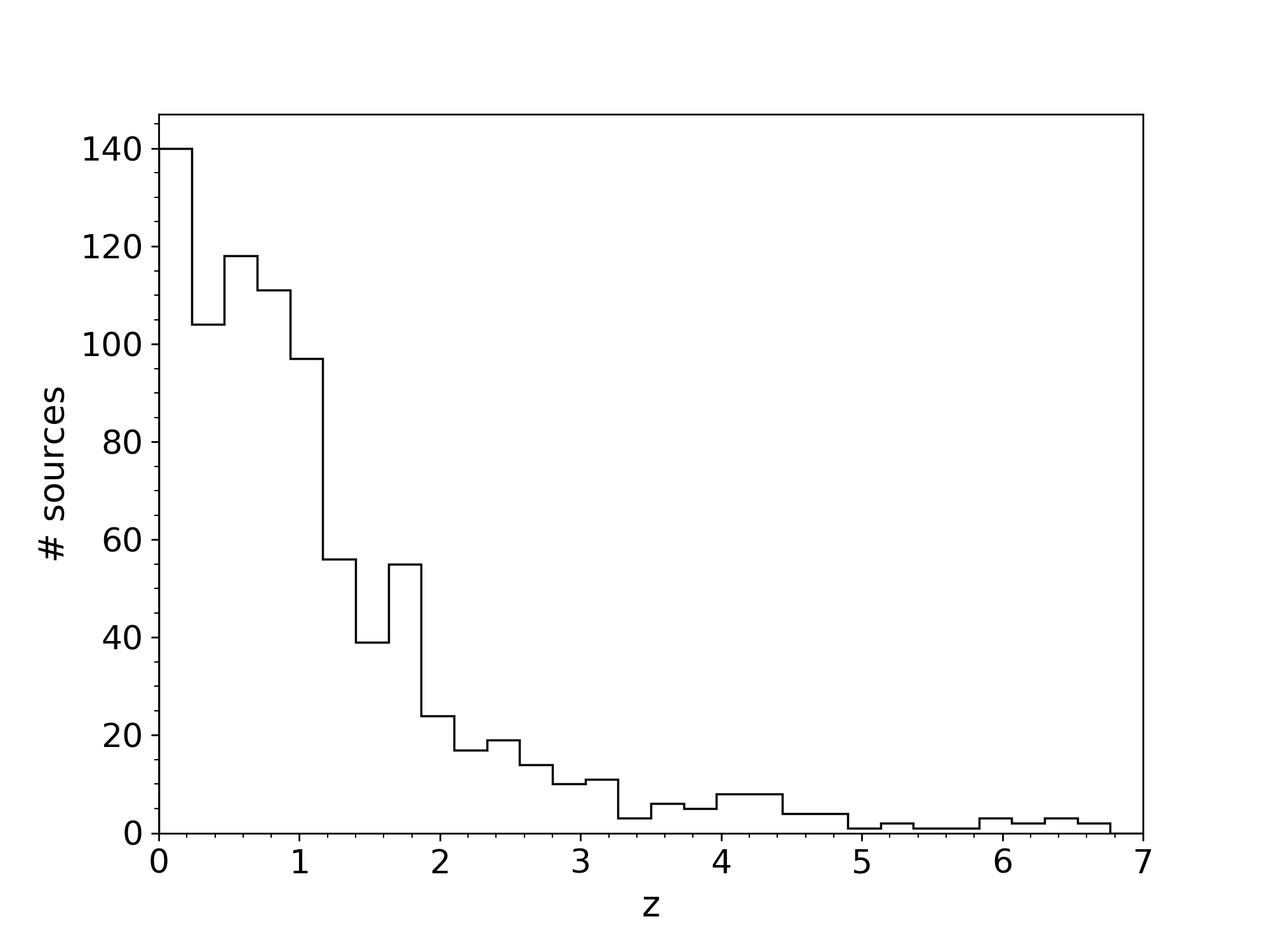}
\includegraphics[width=0.49\textwidth]{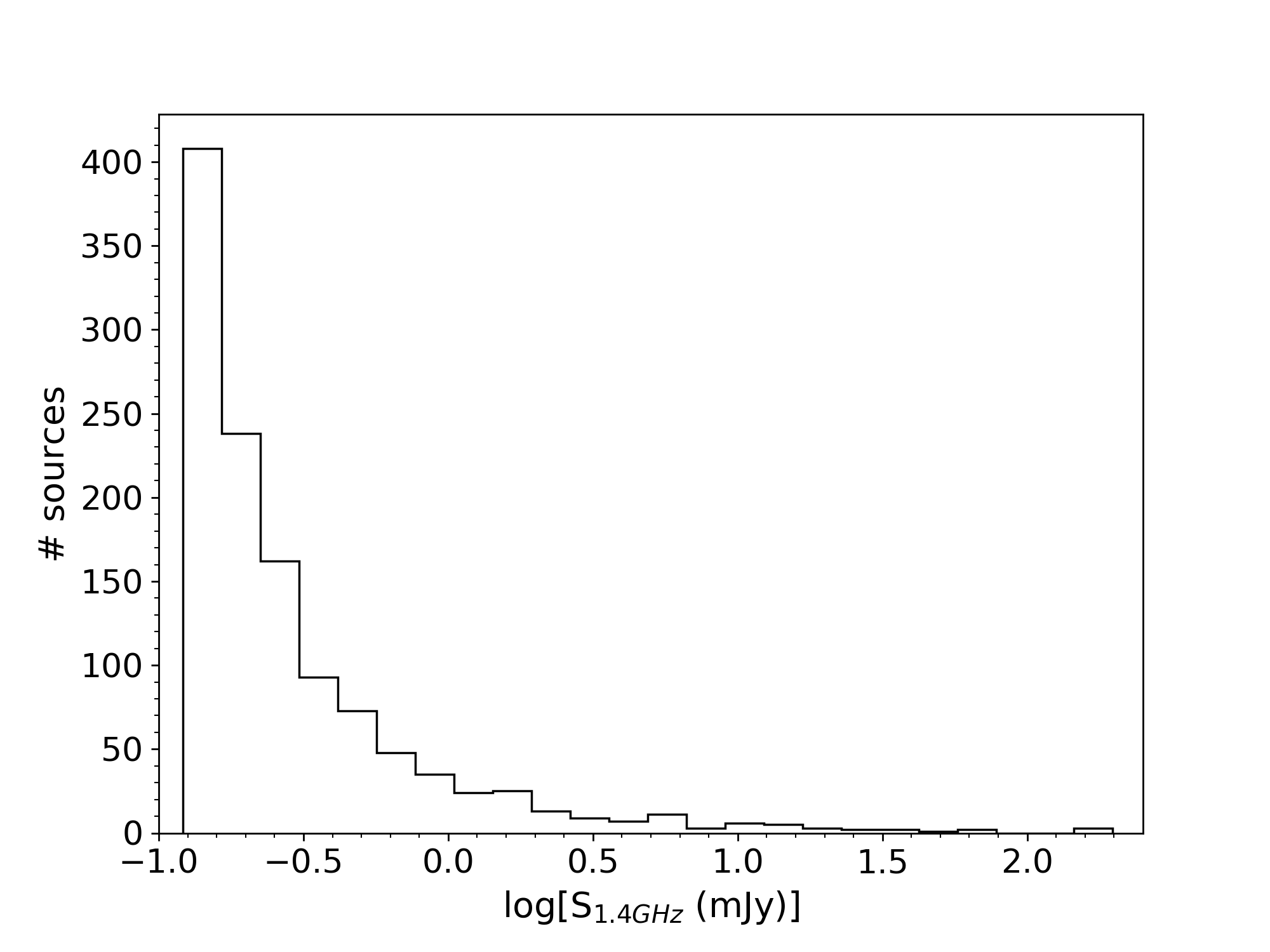}
\caption{Redshift (left) and flux density (right) distributions for our sample.
The redshifts of the IRAC identified sources without spectroscopic measurements or reliable photometric
determinations were estimated using the mean relations with IRAC flux densities shown in
Fig.~\protect\ref{fig:correlations}  (see Sect.\,\ref{sec:data}).}
 \label{fig:distributions}
  \end{center}
\end{figure*}

\subsection{Data}\label{sect:data}


The sample we use is extracted from the full catalogue of radio sources
published in \citet{Prandoni2018}. The sub-region of interest, shown in
Fig.\,\ref{fig:area}, covers $\simeq 1.4\,\hbox{deg}^{2}$ at the center of the
1.4\,GHz WSRT mosaic.  This region is characterized by lowest, roughly uniform
rms noise ($\sim$11\,$\mu$Jy/b) and is completely covered by the
mentioned SERVS Data Fusion and by the UKIDSS-DXS deep K-band mosaic
(\citealt{Lawrence2007}), which were extensively used for the source
identification and classification processes. As described in
\citet{Prandoni2018}, only sources brighter than S$\geq$120\,$\mu$Jy (i.e. with
S/N$\geq$10$\sigma$) were considered. This was motivated by the relatively poor
resolution of the WSRT observations ($\sim$10 arcsec), in combination with the
high number density of the confusion-limited SERVS data set, which prevented
the search for counterparts to be pushed beyond 1.5$-$2 arcsec, to
keep the fraction of misidentifications under control. This means that
the identifications get progressively incomplete going to lower flux densities
(where radio positional errors can be of the order of $\sim$1 arcsec). At
120\,$\mu$Jy the incompleteness was estimated to be $\sim$10$-$15 per cent (see
\citealt{Prandoni2018} for more details).

Table~\ref{tab:sample} reports the number of sources in the full catalogue and
in the region of interest,  as published in \citet{Prandoni2018}. Using
statistical criteria and (radio-only) visual inspection \citet{Prandoni2018}
identified 183 multi-component source in the full catalogue and 45 in the inner
region (all with S$>$120\,$\mu$Jy). These objects were attributed a flux
density equal to the sum of flux densities of the individual components and the
position of the centroid of component positions.

By taking advantage of the available optical/IR images, we further refined this
number by  inspecting radio/optical/IR overlays. After discarding obvious
random pairs the number of confirmed multiple sources decreased to 76,
15 of which are located in the region of interest (see the updated numbers in
Table~\ref{tab:sample}). Correspondingly, the number of
single-component sources with S$>$120\,$\mu$Jy in this region increased
to 1158 so, in total, we have 1173 sources, 921 (908 single and 13 multiple)
of which have an IRAC counterpart within 1.75 arcsec. This corresponds to an
identification rate of $\sim$80\% for the single sources and 87\% for the
multiple sources.

\begin{table}
\centering
\caption{Number of sources in the \citet{Prandoni2018} catalog and in the revised one (used in this paper).}
\begin{tabular}{ccc}
\hline
\hline
Sample & N$_{\rm tot}$ & N$_{\rm multiple}$ \\
\hline
Full catalog (\citealt{Prandoni2018}) & 5997 & 183 \\
Inner S$>$120\,$\mu$Jy & 1110 & 45 \\
\hline
Revised full & 6106 & 76 \\
Inner S$>$120\,$\mu$Jy (IRAC identified) & 1173 (921) & 15 (13) \\
\hline
\hline
\end{tabular}
\label{tab:sample}
\end{table}




Nine hundred sources of our sample have flux density measurements in
the \textit{Spitzer} IRAC band 1, 919 in band 2, 620 in band 3 and 539 in band
4; 502 sources are detected in all the four IRAC bands. Moreover 914 sources
have MIPS 24 $\mu$m fluxes and 829 have $K_{s}$ magnitudes.

Using the recent ``Herschel Extragalactic Legacy Project''
(HELP\footnote{\url{https://herschel.sussex.ac.uk};
\url{https://hedam.lam.fr/HELP/}}; \citealt{Vaccari2016}) catalog
\citep{Shirley2019}, we updated the redshift measurements\footnote{The redshift
estimations provided by \citet{Shirley2019}, computed according to the
\citet{Duncan2018_1, Duncan2018_2} techniques, constitute a refinement and an
update of the measurements included in the SERVS Data Fusion and used by
\citet{Prandoni2018}.} associated to our sources. We got reliable redshift
determinations for 734 of our single sources (196 spectroscopic redshifts plus
538 photometric redshifts based on at least five optical bands). Moreover we
have reliable redshifts for 10 multi-component sources. So, in total, 744
sources ($\sim$63\% of the sample) have reliable redshift determinations.

As shown by Fig.\,\ref{fig:correlations}, IRAC band 1 and band 2 flux densities
are clearly anti-correlated with $\log(z)$\footnote{A relation between
3.6$\mu$m flux density and redshift was also found by
\citet{Orenstein2019}.} (Pearson's correlation coefficients of -0.75 and -0.69,
respectively). The mean relations can be described as
\begin{equation}\label{eq:IRAC}
\log S_{\rm IRAC}= c_{\rm IRAC}-\log z,
\end{equation}
with $c_{\rm IRAC}=1.84$ and a dispersion of 0.35 for band 1 and $c_{\rm
IRAC}=1.79$ with a dispersion of 0.36 for band 2.

For the source classification purposes described in the following sub-sections
we have used a Monte Carlo approach, carrying out 10,000 simulations. In each
simulation the 177 IRAC identified sources without a reliable redshift were attributed a value
of $z$ randomly drawn from a Gaussian distribution with mean given by
eq.~(\ref{eq:IRAC}) for the source $\log S_{\rm IRAC}$ and the associated
dispersion.

We have adopted the mean $z$ and the dispersion from the IRAC band 2 for the 175
(out of 177) sources detected in this band. The other 2 sources were detected
in band 1, and we used the corresponding values.

The \citet{Shirley2019} HELP catalogue also provided SFR determinations (see
\citealt{Malek2018} for the details of the adopted procedure) for 398 sources
($\simeq 34\%$ of the whole sample). SFR estimates for other $\simeq 395$
sources\footnote{The number is slightly different in different simulations,
depending on the redshift assignations.}, $\sim 34\%$ of the sample, were
obtained using the $24\,\mu$m flux density as a SFR indicator, using Fig.~6 of
\citet{Battisti2015}, valid up to $z\simeq 2.95$.

The use of the \citet{Battisti2015} $24\,\mu$m/SFR relation deserves a
comment. Such relation was derived using a sample of SFGs and does not apply if
a substantial fraction of the $24\,\mu$m flux is contributed by an AGN. This
may be a serious limitation for our sample since at its 1.4\,GHz flux density
limit, $\sim 50\%$ of sources are expected to host an AGN
\citep[e.g.,][]{Bonzini2013}. The AGN contribution leads to an overestimate of
the inferred SFR; as a consequence the $L_{\rm 1.4GHz}/\hbox{SFR}$ ratio used
to select RL AGNs (see sub-sect.~\ref{subsec:classification_RLAGN}) is
underestimated and some RL AGNs are missed. However a closer look plays the
problem somewhat down. 

\citet{Elbaz2010} found that also AGNs associated to star forming
galaxies have the same ratios of $L_{\rm IR}$ from $24\,\mu$m to $L_{\rm IR,
tot}(> 30\mu\rm m)$ as SFGs for $\log(L_{\rm IR\,from\, 24\mu m}/L_\odot)\simlt
12.3$. On average, simulations yield IR luminosities above this limit for only
11\% of sources for which we used the \citet{Battisti2015} relation. Of these,
32\% are already classified as RL AGNs. Another 56\%, i.e. 24, show signs of
nuclear activity based on the diagnostics of
sub-sect.~\ref{subsec:classification_SFG_RQAGN}; their SFRs might have been
overestimated to the point that they are all misclassified RL AGNs. We have
conservatively added the maximum number of misclassifications (24) to the error
on the classification of RL and RQ AGNs. As we will see in the following
sub-sections, this turns out to be a sub-dominant contribution. 

As a counter-check we have assumed that 50\% of the $24\,\mu$m flux of
sources for which we used the \citet{Battisti2015} relation and were classified
as RQ AGNs is of nuclear origin. Decreasing by a factor of 2 their SFRs,
simulations give that, on average, 21 of them are reclassified as RL AGNs. This
confirms that our previous estimate of the classification uncertainty is
conservative.


The redshift and 1.4\,GHz flux density distributions of our sample are shown in
Fig.\,\ref{fig:distributions}.  The median values are $z\simeq 0.85$ and
$S_{1.4\,\rm GHz}\simeq 0.208\,$mJy. 


\subsection{RL AGN classification}\label{subsec:classification_RLAGN}

\begin{figure}
\begin{center}
\includegraphics[width=0.49\textwidth]{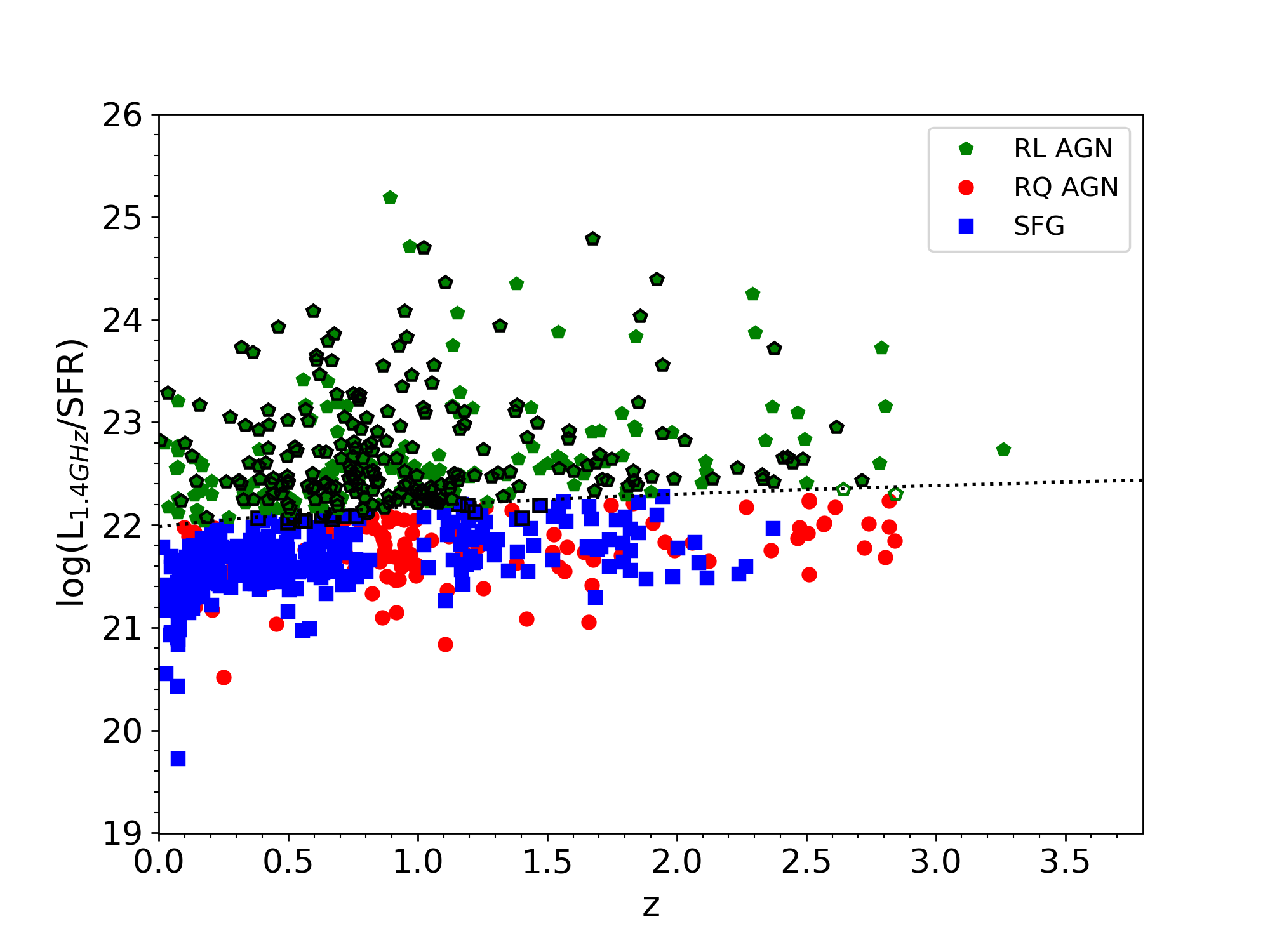}
\caption{Radio-loudness diagnostics, $\log(L_{\rm 1.4GHz}/\hbox{SFR})$, versus $z$.
The dotted black line represents the redshift-dependent (3$\sigma$) threshold given by \citet{Smolcic2017b}:
$\log(L_{\rm 1.4GHz}/\hbox{SFR})=21.984\times(1+z)^{0.013}$, with $L_{\rm1.4GHz}$ in $\hbox{W}\,\hbox{Hz}^{-1}$
and SFR in $M_{\odot}\,\hbox{yr}^{-1}$. Pentagons, squares and circles are RL AGNs, SFGs
and RQ AGNs, respectively. Filled pentagons refer to RL AGNs classified with
$\log(L_{\rm 1.4GHz}/\hbox{SFR})$ above this threshold.
Symbols with thick black borders represent sources for which no estimate of the SFR is available
from the HELP catalogue and also miss the MIPS $24\,\mu$m flux density from which it could be derived.
These objects were classified on the basis of their $5\,\sigma$ upper limit at $24\,\mu$m (through the \citealt{Battisti2015}
relation, see sub-sect.\,\ref{sect:data}). Whenever such upper limit was absent or
did not allow a reliable classification we adopted the criterion by
\citet[][their eq.\,(1)]{Magliocchetti2017} which classified them as RL AGNs; these objects are
represented by open pentagons (note that most of the RL AGNs classified through the
\citealt{Magliocchetti2017} criterion don't have SFR estimates or upper limits
and therefore are not present in this plot).
 }
 \label{fig:RLAGN_D1}
  \end{center}
\end{figure}

\begin{figure}
\begin{center}
\includegraphics[width=0.49\textwidth]{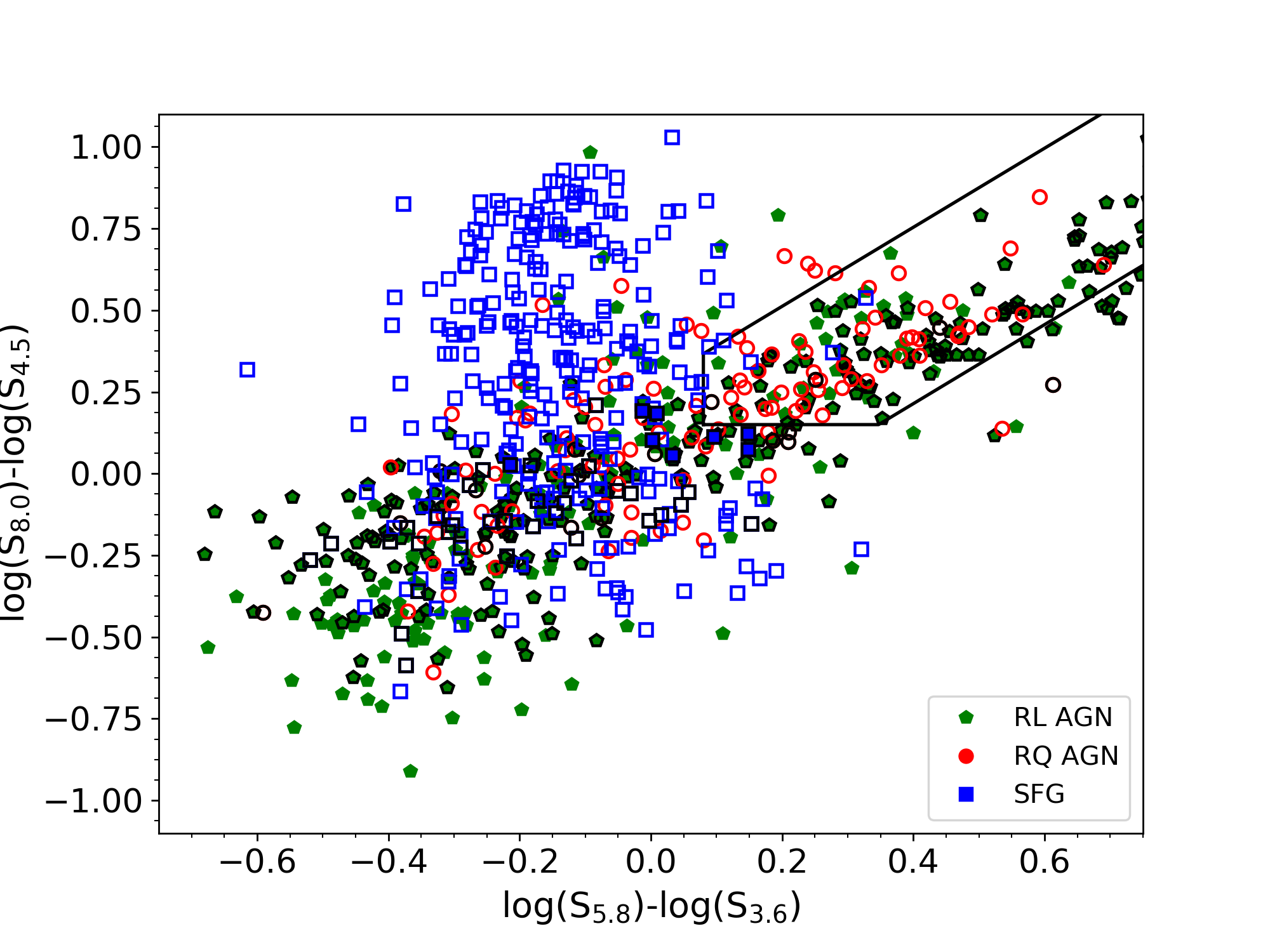}
\caption{IRAC colour - colour diagram: $\log({S_{8.0\,\mu\rm m}})-\log({S_{4.5\,\mu\rm m}})$ vs.
$\log({S_{5.8\,\mu\rm m}})-\log({S_{3.6\,\mu\rm m}})$ used to divide the non-RL AGNs between RQ AGNs and SFGs
(see sub-sect.\,\ref{subsec:classification_SFG_RQAGN}).
The ``RQ AGN region'' (bounded by the solid black line) is defined by:
$\log{[S_{8.0\,\mu\rm m}/S_{4.5\,\mu\rm m}]} \geq 0.15$; $\log{[S_{5.8\,\mu\rm m}/S_{3.6\,\mu\rm m}]} \geq 0.08$;
$\log{[S_{8.0\,\mu\rm m}/S_{4.5\,\mu\rm m}]} \geq 1.21\times\log{[S_{5.8\,\mu\rm m}/S_{3.6\,\mu\rm m}]} - 0.27$;
$\log{[S_{8.0\,\mu\rm m}/S_{4.5\,\mu\rm m}]} \leq 1.21\times\log{[S_{5.8\,\mu\rm m}/S_{3.6\,\mu\rm m}]}+0.27$ (\citealt{Donley2012}).
Circles, squares and pentagons represent RQ AGNs, SFGs and RL AGNs, respectively. The filled/open symbols refer to objects
classified using this/another diagnostic or criterion (see text). Symbols with heavy black borders correspond to objects for which
not all IRAC flux densities are available. The missing flux densities were replaced with their $5\,\sigma$
upper limits. }
 \label{fig:RLAGN_D2}
  \end{center}
\end{figure}


The multi-component sources were classified as RL AGNs. To select RL AGNs among
the single-component sources  we used two criteria. Whenever an estimate of the
SFR was available (either from the HELP catalogue or derived from the
$24\,\mu$m data) we classified as RL AGNs sources with radio luminosity to SFR
ratios above the redshift-dependent threshold $\log(L_{\rm
1.4GHz}/\hbox{SFR})=21.984\times(1+z)^{0.013}$, with $L_{\rm1.4GHz}$ in
$\hbox{W}\,\hbox{Hz}^{-1}$ and SFR in $M_{\odot}\,\hbox{yr}^{-1}$
\citep{Smolcic2017b}. This criterion is illustrated by Fig.~\ref{fig:RLAGN_D1}.

For the other sources we adopted the \citet{Magliocchetti2017} criterion (their
eq.\,1), i.e. we classified as RL AGNs sources whose radio luminosity exceeds
\begin{equation}\label{eq:Maglio}
\log L_{\rm cross}(z)= \left\{ \begin{array}{r@{\quad}l} 22.8 +z & \hbox{for}\  z  \le 1.8 \\
                        24.6 & \hbox{for}\  z  > 1.8 \end{array} \right.
\end{equation}
Note that here radio luminosities are in $\hbox{W}\,\hbox{Hz}^{-1}$ while in
\citet{Magliocchetti2017} they are in
$\hbox{W}\,\hbox{Hz}^{-1}\,\hbox{sr}^{-1}$.

Using these criteria we identified 458 sources as RL AGNs, most of them
($\simeq 80\%$) through the first criterion. Adding the 15 multi-component
sources we end up with a total of 473 RL AGNs.

\subsection{SFG/RQ AGN classification}\label{subsec:classification_SFG_RQAGN}

\begin{figure}
\begin{center}
\includegraphics[width=0.49\textwidth]{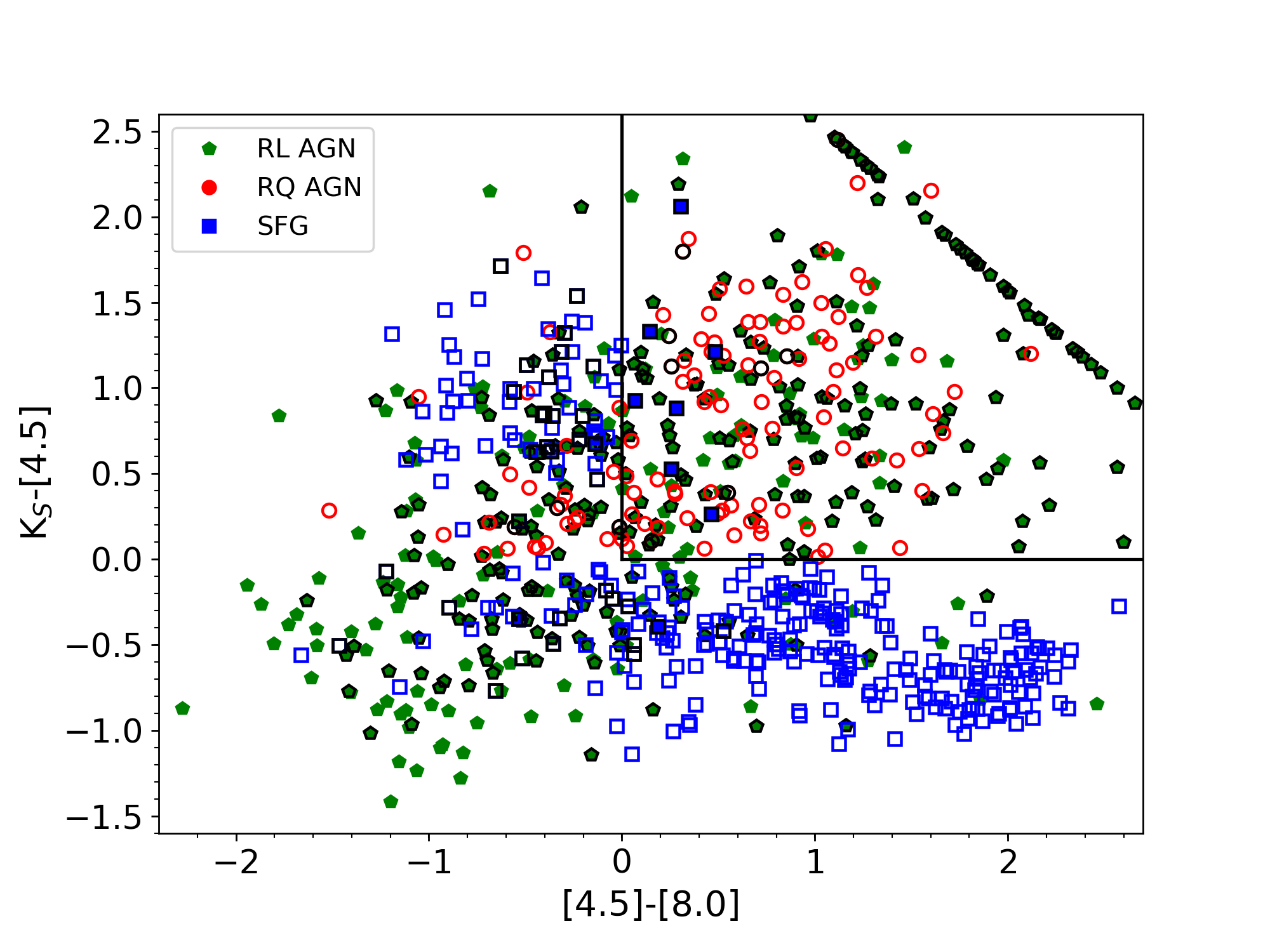}
\caption{$K_s$--IRAC (KI) diagnostic plot. The $K_{s}$--[4.5] vs. [4.5]--[8.0] colours,
with magnitudes converted to the AB system, are used to classify the non-RL
AGN sources as either RQ AGNs or SFGs
(see sub-sect.\,\ref{subsec:classification_SFG_RQAGN}). The ``RQ AGN region"
(bounded by the solid black line) is defined by: $K_{s}-$[4.5]$>0$; [4.5]$-$[8.0]$>0$
(\citealt{Messias2012}). Circles, squares and pentagons represent RQ AGNs, SFGs and RL AGNs,
respectively. Filled and open symbols represent sources classified with this or other diagnostics/criteria,
respectively. Symbols with thick black borders refer to sources undetected in at least one
of the three bands involved in this diagnostic, The missing magnitudes have been replaced
by the corresponding $5\,\sigma$ detection limits. }
 \label{fig:KI}
  \end{center}
\end{figure}

The 450 non-RL AGN (IRAC identified) sources were considered as either SFGs or RQ AGNs. To
discriminate among the two populations, we adopted the redshift-dependent
criterion discussed by \citet{Messias2012}, that involves the $K_{s}$, 4.5, 8.0
and 24\,$\mu$m AB magnitudes\footnote{$K_{s}$ Vega magnitudes were converted
into AB magnitudes as $K_{s, \rm AB}= K_{s, \rm Vega}+1.85$.  Flux densities
(in $\mu$Jy) were converted into AB magnitudes through the relation:
$\hbox{ABmag}=-2.5\times \log (S_{\nu}\times 10^{-29})-48.6$. }. We classified
sources as RQ AGNs when:
\begin{equation}
\begin{array}[c]{ll}
K_{s} - [4.5] > 0\, & {\rm for}\,z \leq 1; \\\relax
[4.5] - [8.0] > 0\, & {\rm for}\,1 < z \leq 2.5; \\\relax
[8.0] - [24] > 1\,  & {\rm for}\,z > 2.5.\\
\end{array}
\end{equation}
Sources missing either the $K_s$ (for z$\leq$1) or the $24\,\mu$m (for $z>
2.5$) magnitudes  were classified adopting either the IRAC colour-colour
diagram ($\log{S_{8.0\,\mu\rm m}} - \log{S_{4.5\,\mu\rm m}}$ vs.
$\log{S_{5.8\,\mu\rm m}} - \log{S_{3.6\,\mu\rm m}}$; Fig.~\ref{fig:RLAGN_D2})
or the $K_s$--IRAC (KI) diagnostic plot ($K_{s}$--[4.5] vs. [4.5]--[8.0];
Fig.~\ref{fig:KI}).

In the IRAC colour-colour diagram, the ``RQ AGN region'' is defined  by
\citep{Donley2012}:
\begin{equation}
\begin{array}[c]{l}
\log{[S_{8.0\,\mu\rm m}/S_{4.5\,\mu\rm m}]} \geq 0.15;\\\relax
\log{[S_{5.8\,\mu\rm m}/S_{3.6\,\mu\rm m}]} \geq 0.08;\\\relax
\log{[S_{8.0\,\mu\rm m}/S_{4.5\,\mu\rm m}]} \geq 1.21\times\log{[S_{5.8\,\mu\rm m}/S_{3.6\,\mu\rm m}]} - 0.27;\\\relax
\log{[S_{8.0\,\mu\rm m}/S_{4.5\,\mu\rm m}]} \leq 1.21\times\log{[S_{5.8\,\mu\rm m}/S_{3.6\,\mu\rm m}]}+0.27.\\
\end{array}
\end{equation}
In the KI diagnostic plot, the ``RQ AGN region'' is defined  by
\citep{Messias2012}:
\begin{equation}
\begin{array}[c]{l}
K_{s} - [4.5]>0; \\\relax
[4.5] - [8.0]>0. \\
\end{array}
\end{equation}
%
%
%
None of these diagnostics could be applied to only 16 (IRAC identified)
sources, which remained unclassified\footnote{The total number of unclassified
sources, including the IRAC unidentified single-component sources (for which
the application of our diagnostics was not possible as well), is 266 (i.e.
16+250). Actually, the total number of IRAC unidentified sources is
$1173-921=252$ (see sub-sect.~\ref{sect:data}. Two of them, however, are
multi-component and were therefore classified as RL AGNs, as stated in
sub-sect.~\ref{subsec:classification_RLAGN}.)}. The IRAC and the KI
diagnostics could be simultaneously applied only to very few sources ($\leq3$,
the number varied in the different simulations\footnote{The simulations, based
on a Monte Carlo approach, are presented in sub-sect.\,\ref{sect:data}.}) and
the classification always agreed.

\begin{table*}
\caption{Sample of the catalogue content. The complete catalogue
is available as supplementary material in the electronic version of the paper.}
\centering
\begin{tabular}{lccccccc}
\hline
IAU name & RA J2000 & DEC J2000 & Classification & Class. rel. & z & S$_{\rm1.4GHz}$ & $\sigma$ \\
& ${\rm [deg]}$ & ${\rm [deg]}$ &  & \% & & ${\rm [mJy]}$ & ${\rm [mJy]}$\\
\hline
LHWJ105237+573101  &  163.1549  &  57.5170  &  RL AGN  &  100  &  0.71  &  59.69  &  0.012  \\
LHWJ105150+573244  &  162.9603  &  57.5456  &  RL AGN  &  100  &  0.19  &  16.18  &  0.012  \\
LHWJ105427+573646  &  163.6127  &  57.6130  &  RL AGN  &  100  &  0.32  &  180.64  &  0.014  \\
LHWJ104846+573750  &  162.1955  &  57.6308  &  RL AGN  &  100  &  0.92  &  0.49  &  0.011  \\
LHWJ105337+574242  &  163.4060  &  57.7118  &  RL AGN  &  100  &  0.66  &  2.08  &  0.011  \\
LHWJ105326+574544  &  163.3614  &  57.7623  &  RL AGN  &  100  &  0.87  &  7.02  &  0.010  \\
LHWJ105243+574813  &  163.1810  &  57.8037  &  RL AGN  &  100  &  1.03  &  1.86  &  0.010  \\
LHWJ105412+575651  &  163.5536  &  57.9476  &  RL AGN  &  100  &  0.64  &  2.86  &  0.010  \\
LHWJ105027+581517  &  162.6132  &  58.2548  &  RL AGN  &  100  &  0.50  &  2.02  &  0.010  \\
LHWJ105342+582108  &  163.4258  &  58.3525  &  RL AGN  &  100  &  -  &  0.93  &  0.010  \\
LHWJ105351+582519  &  163.4656  &  58.4222  &  RL AGN  &  100  &  0.80  &  58.50  &  0.011  \\
LHWJ105532+583156  &  163.8854  &  58.5323  &  RL AGN  &  100  &  -  &  1.75  &  0.011  \\
LHWJ105116+583829  &  162.8190  &  58.6414  &  RL AGN  &  100  &  -  &  1.45  &  0.012  \\
LHWJ105314+572411  &  163.3099  &  57.4032  &  SFG  &  100  &  1.69  &  0.12  &  0.011  \\
LHWJ105217+572127  &  163.0735  &  57.3576  &  RQ AGN  &  100  &  0.05  &  0.21  &  0.011  \\
LHWJ105237+572148  &  163.1559  &  57.3635  &  RQ AGN  &  100  &  0.48  &  0.16  &  0.011  \\
LHWJ105308+572222  &  163.2835  &  57.3730  &  RL AGN  &  100  &  0.39  &  0.30  &  0.011  \\
LHWJ104946+575140  &  162.4438  &  57.8612  &  RL AGN  &  100  &  0.04  &  0.16  &  0.010  \\
LHWJ105120+572253  &  162.8347  &  57.3815  &  RL AGN  &  100  &  1.22  &  0.28  &  0.011  \\
LHWJ105241+572320  &  163.1727  &  57.3890  &  RL AGN  &  100  &  1.06  &  1.86  &  0.011  \\
LHWJ105158+572329  &  162.9955  &  57.3916  &  RL AGN  &  100  &  -  &  0.24  &  0.011  \\
LHWJ105305+572330  &  163.2721  &  57.3919  &  RQ AGN  &  100  &  0.99  &  0.15  &  0.011  \\
LHWJ105254+572341  &  163.2261  &  57.3950  &  RL AGN  &  100  &  0.76  &  0.20  &  0.011  \\
LHWJ105356+572355  &  163.4837  &  57.3986  &  RQ AGN  &  100  &  1.35  &  0.13  &  0.011  \\
LHWJ105348+572358  &  163.4520  &  57.3996  &  SFG  &  100  &  1.67  &  0.12  &  0.011  \\
LHWJ105031+572401  &  162.6314  &  57.4003  &  RL AGN  &  100  &  -  &  0.13  &  0.012  \\
LHWJ105406+572413  &  163.5254  &  57.4039  &  RQ AGN  &  100  &  0.77  &  0.16  &  0.012  \\
LHWJ105053+572427  &  162.7223  &  57.4076  &  SFG  &  100  &  0.40  &  0.31  &  0.012  \\
LHWJ105239+572432  &  163.1651  &  57.4089  &  RQ AGN  &  100  &  1.11  &  0.16  &  0.011  \\
LHWJ105023+572439  &  162.5971  &  57.4108  &  SFG  &  100  &  1.19  &  0.29  &  0.011  \\
LHWJ105242+572444  &  163.1765  &  57.4124  &  SFG  &  100  &  0.08  &  0.27  &  0.011  \\
LHWJ105314+572449  &  163.3085  &  57.4138  &  RL AGN  &  100  &  0.69  &  0.17  &  0.011  \\
LHWJ105212+572453  &  163.0522  &  57.4148  &  SFG  &  100  &  0.50  &  0.23  &  0.011  \\
LHWJ105301+572521  &  163.2570  &  57.4227  &  RL AGN  &  100  &  0.57  &  0.29  &  0.011  \\
LHWJ105008+572514  &  162.5349  &  57.4207  &  RQ AGN  &  100  &  2.51  &  0.16  &  0.011  \\
LHWJ105356+572516  &  163.4840  &  57.4211  &  SFG  &  100  &  0.50  &  0.12  &  0.011  \\
LHWJ105343+572531  &  163.4314  &  57.4253  &  RL AGN  &  100  &  0.75  &  0.31  &  0.011  \\
LHWJ105049+572527  &  162.7054  &  57.4243  &  SFG  &  100  &  0.37  &  0.14  &  0.011  \\
LHWJ105035+572532  &  162.6460  &  57.4258  &  RL AGN  &  94  &  -  &  0.20  &  0.011  \\
LHWJ105421+572544  &  163.5883  &  57.4290  &  RQ AGN  &  100  &  0.20  &  1.06  &  0.012  \\
LHWJ105232+572543  &  163.1370  &  57.4286  &  RQ AGN  &  100  &  0.56  &  0.14  &  0.011  \\
LHWJ105403+572553  &  163.5159  &  57.4314  &  RQ AGN  &  100  &  2.82  &  0.18  &  0.012  \\
LHWJ105359+572559  &  163.4995  &  57.4333  &  RQ AGN  &  100  &  2.82  &  0.24  &  0.012  \\
LHWJ105035+572607  &  162.6490  &  57.4355  &  SFG  &  100  &  0.36  &  0.17  &  0.011  \\
LHWJ105149+572635  &  162.9582  &  57.4432  &  RL AGN  &  100  &  1.14  &  0.27  &  0.010  \\
LHWJ105032+572646  &  162.6362  &  57.4463  &  RL AGN  &  100  &  -  &  0.62  &  0.011  \\
LHWJ105113+572654  &  162.8060  &  57.4484  &  RQ AGN  &  100  &  0.63  &  0.29  &  0.010  \\
LHWJ105351+572700  &  163.4662  &  57.4501  &  RQ AGN  &  100  &  0.92  &  0.21  &  0.011  \\
LHWJ105204+572657  &  163.0170  &  57.4494  &  x  &  0  &  1.48  &  0.15  &  0.011  \\
LHWJ105104+572719  &  162.7707  &  57.4553  &  RL AGN  &  75  &  -  &  0.20  &  0.010  \\
... & ... & ... & ... & ... & ... & ... & ... \\
\hline
\multicolumn{8}{|p{\linewidth}|}{\textit{Notes}.}\\
\multicolumn{8}{|p{\linewidth}|}{1. In the case of extended sources, S$_{\rm 1.4GHz}$ is the flux density measured by
summing all the pixels above a $3\,\sigma$ threshold (see \citealt{Prandoni2018} for details).}\\
\multicolumn{8}{|p{\linewidth}|}{2. $\sigma$ is the local rms noise value.}\\
\multicolumn{8}{|p{\linewidth}|}{3. For IRAC identified sources lacking a reliable redshift determination we give
the reliability of the classification (Class. rel.) defined as the percentage of simulations
described in sub-sect.~\ref{sect:data} yielding such classification.}\\
\multicolumn{8}{|p{\linewidth}|}{4. The X-ray analysis (see sub-sect. \ref{subsec:classification_SFG_RQAGN})
 shows that the number of SFGs should be corrected by a factor of $\sim$0.88. However, this correction has only
 a statistical meaning and cannot be done source by source. In the table we have
 corrected the classification (from SFG to RQ AGN) of only the three sources in the XMM-\textit{Newton}
 survey with $L_{X}>10^{42}\,\hbox{erg}\,\hbox{s}^{-1}$.} \\
\end{tabular}
\label{tab:fluxes}
\end{table*}

A check of our classification was performed taking advantage of the
XMM-\textit{Newton} survey of the LH field \citep{Brunner2008} which covers a
small fraction ($\sim 10\%$) of our field. In the overlapping area there are
121 of our sources, 46 classified as RL AGNs, 24 as SFGs, 24 as RQ AGNs and 27
unclassified. Using a $4''$ search radius (the XMM-\textit{Newton} positional
errors are generally $<1''$ but in some cases reach $2.3''$), we found 32
matches: 11 classified as RL AGNs, 9 as RQ AGNs, 8 as SFGs and 4 unclassified.
Three ``SFGs'' turned out to have 2--10\,keV luminosities
$L_X>10^{42}\,\hbox{erg}\,\hbox{s}^{-1}$. The X-ray luminosity associated to
star-formation rarely exceeds $L_X=10^{41.5}\,\hbox{erg}\,\hbox{s}^{-1}$
\citep{Ranalli2003, Lehmer2010}; hence such sources are most likely
misclassified RQ AGNs. The other 5 have $L_X$ in the range 7$\times$10$^{38}$
$-$ 4$\times$10$^{41}$erg/s, hence compatible with the SFG classification,
although the presence of a low-luminosity AGN cannot be excluded.

Based on this result, we have corrected the number of SFGs by a factor
$(24-3)/24\simeq0.88$, thus decreasing it from 310 to 271. Since the radio-loudness
diagnostics indicate that these X-ray sources are not RL, we reclassify them as
RQ AGNs. If we conservatively consider as uncertain the classification of the 5
sources with X-ray luminosity compatible with the SFG classification, the
contribution to the error budget of the classification uncertainty is
$5/(24-3)\simeq 24\%$. After this correction the result of the classification
is $271\pm 74$ SFGs and $163 \pm 74$ RQ AGNs.

Note that the contributions to the global error of uncertainties
associated to the $24\,\mu$m--SFR conversion is sub-dominant. 

%

The source catalogue, including our classification, is available as
supplementary material in the electronic version of the paper. The catalogue
gives the IAU name, the equatorial coordinates (J2000), the source
classification, its redshift (if available), the 1.4\,GHz flux density and the
rms. An example of the content of the catalogue is given in
Table~\ref{tab:fluxes}.


\begin{figure*}
\begin{center}
\includegraphics[width=0.49\textwidth]{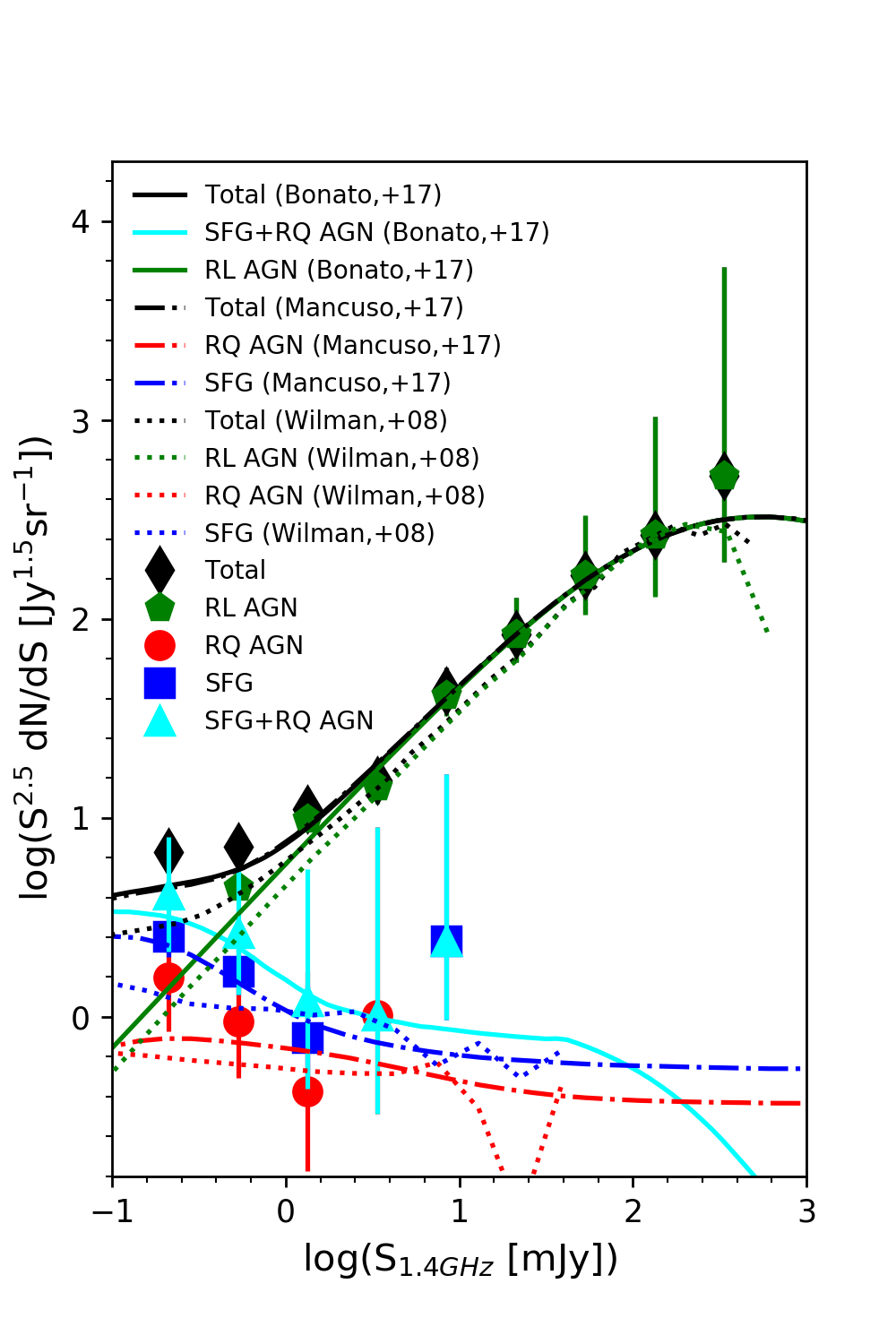}
\includegraphics[width=0.49\textwidth]{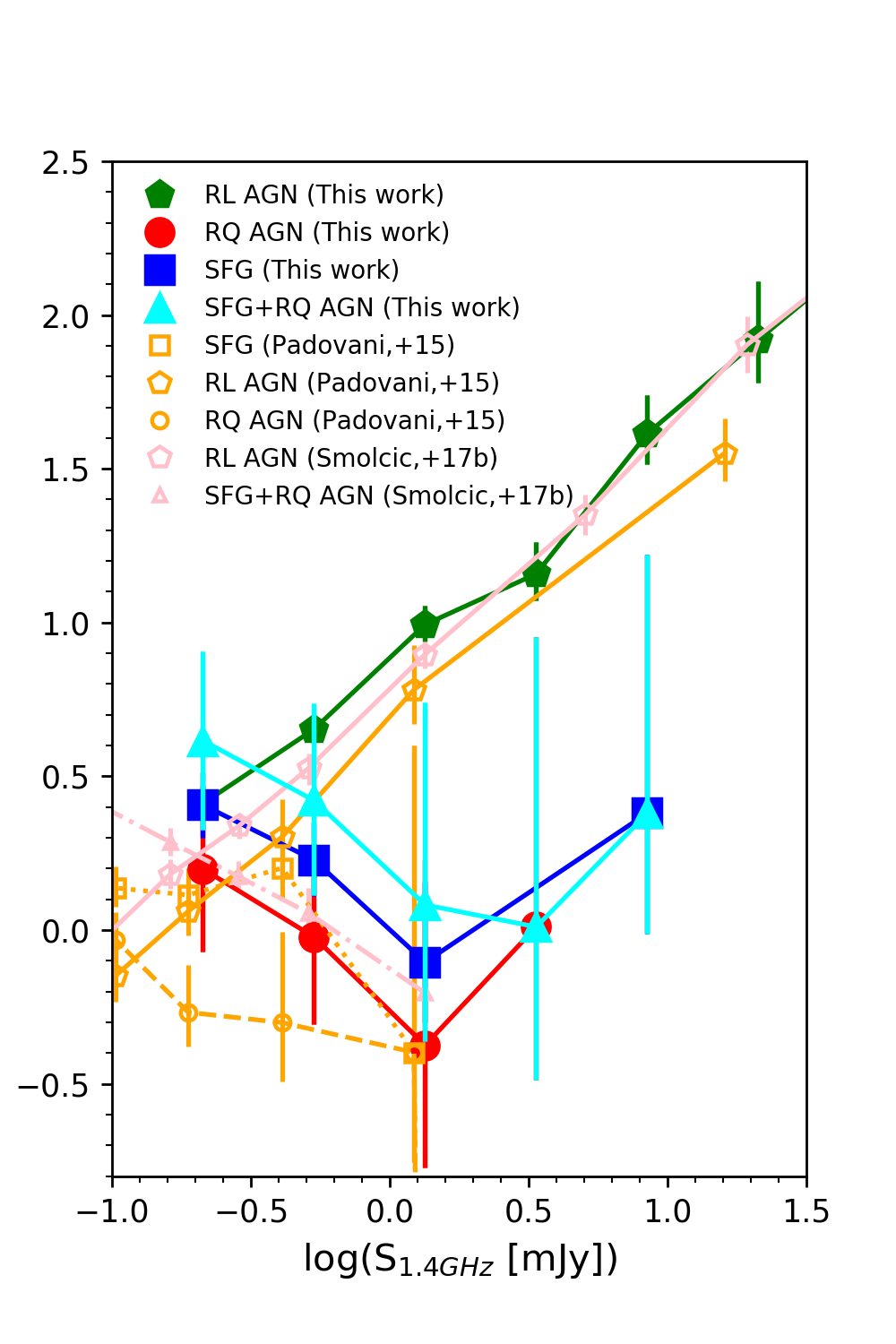}
\caption{Comparison  of our estimate of the number counts at 1.4\,GHz of RL AGNs, RQ AGNs, SFGs, RQ AGNs+SFGs
and total, with models and with other estimates (left and right panel, respectively).
Models are from \citet[][RL AGNs, SFGs+RQ AGNs and total]{Bonato2017},
\citet[][SFGs, RQ AGNs and total]{Mancuso2017} and \citet[][RL AGNs, SFGs, RQ AGNs and total]{Wilman2008}.
The Tiered Radio Extragalactic Continuum Simulation (T-RECS) counts \citet{Bonaldi2019}, not shown,
are very close to the \citet{Bonato2017} ones.
The observational estimates are from \citet{Padovani2015}, \citet{Smolcic2017b} and from the present work.
The lines connecting the points (on the right panel) are only meant to guide the eye. }
 \label{fig:nc_model}
  \end{center}
\end{figure*}

\section{Results}\label{sec:results}



\subsection{Number counts}\label{subsec:nc}

Our estimates of the Euclidean normalized differential number counts of RL
AGNs, RQ AGNs and SFGs, and the total are reported in Tab.\,\ref{tab:counts}
and displayed in Fig.\,\ref{fig:nc_model}. The unclassified and/or unidentified sources have been distributed among the three
populations in proportion to their numbers in each flux density bin. 

The number counts of SFGs and of RQ AGNs were corrected for misclassifications
based on the X-ray data, as described above.  The uncertainties on the number
counts are the sum in quadrature of Poisson errors and of the variance in the
simulations. In the case of SFGs and RQ AGNs, we have also taken into account
the conservative classification uncertainty
(sub-sect.~\ref{subsec:classification_SFG_RQAGN}).

Obviously the adoption of the same correction factor for all flux
density bins is a rough fix, but it is the best we can do since the very poor
statistics prevents any possibility of a more refined, and realistic,
treatment. This is taken into account applying a conservative error estimate,
including the classification uncertainty. 

In the left panel of Fig.\,\ref{fig:nc_model} our estimates are compared with
the models by \citet{Bonato2017}, \citet{Mancuso2017} and \citet{Wilman2008}.
The models by \citet{Wilman2008} and by \citet{Bonato2017} provide a good fit
of our observational determination of the counts of RL AGNs. The model by
\citet{Bonato2017}, which does not distinguish between SFGs and RQ AGNs,
reproduces quite well also their summed counts. \citet{Wilman2008} slightly
under-predict the counts of both RQ AGNs and SFGs, and of the total population
as well, as noted also by \citet{Prandoni2018}. The model by
\citet{Mancuso2017} nicely
agrees with our counts.

As illustrated by the right-hand panel of Fig.\,\ref{fig:nc_model}, our
estimates of the counts of each population agree, within the errors, with those
by \citet{Padovani2015} and \citet{Smolcic2017b} who used different samples and
different approaches. These studies converge in indicating that RL AGNs
dominate the 1.4\,GHz counts above 200--$300\,\mu$Jy, while SFGs$+$RQ AGNs take
over at fainter flux densities.
The counts of RQ AGNs are approximately
parallel to those of SFGs, indicating similar evolution, as previously pointed
out by \citet{Padovani2015}. Our estimates give surface densities of RQ AGNs
only slightly below those of SFGs, at variance with \citet{Padovani2015} who
found abundances of RQ AGNs lower than those of SFGs by factors of 2 to 3.

\begin{table*}
\centering
\caption{Estimates of the 1.4\,GHz Euclidean normalized differential
counts, $S^{2.5}dN/dS [{\rm Jy}^{1.5}{\rm sr}^{-1}]$, for the RL AGN,
SFG and RQ AGN populations, and total. On the right of the counts
we give the number, $N$, of sources in each bin. It includes the contributions of unidentified
and unclassified sources as well as the correction for the misclassification indicated
by X-ray data (for SFGs and RQ AGNs). The errors are dominated by classification uncertainties. 
See sub-sect.\,\ref{subsec:nc}.}
\begin{tabular}{ccccccccc}
\hline
\hline
$\log S$ & $\log({\rm Counts}_{\rm TOT})$ & ${\rm N}_{\rm TOT}$ & $\log({\rm Counts}_{\rm RL\, AGN})$ & ${\rm N}_{\rm RL\, AGN}$ & $\log({\rm Counts}_{\rm SFG})$ & ${\rm N}_{\rm SFG}$ & $\log({\rm Counts}_{\rm RQ\, AGN})$ & ${\rm N}_{\rm RQ\, AGN}$ \\
$\rm[mJy]$ & $[{\rm Jy}^{1.5}{\rm sr}^{-1}]$ & & $[{\rm Jy}^{1.5}{\rm sr}^{-1}]$ & & $[{\rm Jy}^{1.5}{\rm sr}^{-1}]$ & & $[{\rm Jy}^{1.5}{\rm sr}^{-1}]$ & \\
\hline
-0.67  & $ 0.83 ^{+ 0.02 }_{- 0.02 }$ &  805  & $ 0.41 ^{+ 0.03 }_{- 0.03 }$ &  309  & $ 0.41 ^{+ 0.11 }_{- 0.11 }$ &  306  & $ 0.20 ^{+ 0.27 }_{- 0.27 }$ & 190  \\
-0.27  & $ 0.85 ^{+ 0.03 }_{- 0.03 }$ &  216  & $ 0.65 ^{+ 0.05 }_{- 0.04 }$ &  136  & $ 0.23 ^{+ 0.12 }_{- 0.12 }$ &  51  & $ -0.02 ^{+ 0.29 }_{- 0.28 }$ & 29  \\
0.13  & $ 1.04 ^{+ 0.05 }_{- 0.05 }$ &  84  & $ 0.99 ^{+ 0.06 }_{- 0.05 }$ &  75  & $ -0.10 ^{+ 0.26 }_{- 0.20 }$ &  6  & $ -0.37 ^{+ 0.60 }_{- 0.40 }$ & 3  \\
0.53  & $ 1.19 ^{+ 0.10 }_{- 0.08 }$ &  29  & $ 1.16 ^{+ 0.10 }_{- 0.09 }$ &  27  & $-$ & $-$  & $ 0.01 ^{+ 0.94 }_{- 0.50 }$ & 2  \\
0.93  & $ 1.64 ^{+ 0.12 }_{- 0.09 }$ &  21  & $ 1.61 ^{+ 0.12 }_{- 0.10 }$ &  20  & $ 0.38 ^{+ 0.84 }_{- 0.39 }$ &  1  & $-$ & $-$  \\
1.33  & $ 1.92 ^{+ 0.19 }_{- 0.14 }$ &  10  & $ 1.92 ^{+ 0.19 }_{- 0.14 }$ &  10  & $-$ & $-$ & $-$ & $-$  \\
1.73  & $ 2.22 ^{+ 0.30 }_{- 0.19 }$ &  5  & $ 2.22 ^{+ 0.30 }_{- 0.20 }$ &  5  & $-$ & $-$ & $-$ & $-$  \\
2.13  & $ 2.42 ^{+ 0.59 }_{- 0.31 }$ &  2  & $ 2.42 ^{+ 0.59 }_{- 0.31 }$ &  2  & $-$ & $-$ & $-$ & $-$  \\
2.53  & $ 2.72 ^{+ 1.05 }_{- 0.43 }$ &  1  & $ 2.72 ^{+ 1.05 }_{- 0.43 }$ &  1  & $-$ & $-$ & $-$ & $-$  \\
\hline
\hline
\end{tabular}
\label{tab:counts}
\end{table*}

\subsection{Luminosity functions}\label{subsec:LF}
\begin{figure*}
\begin{center}
\includegraphics[width=0.49\textwidth]{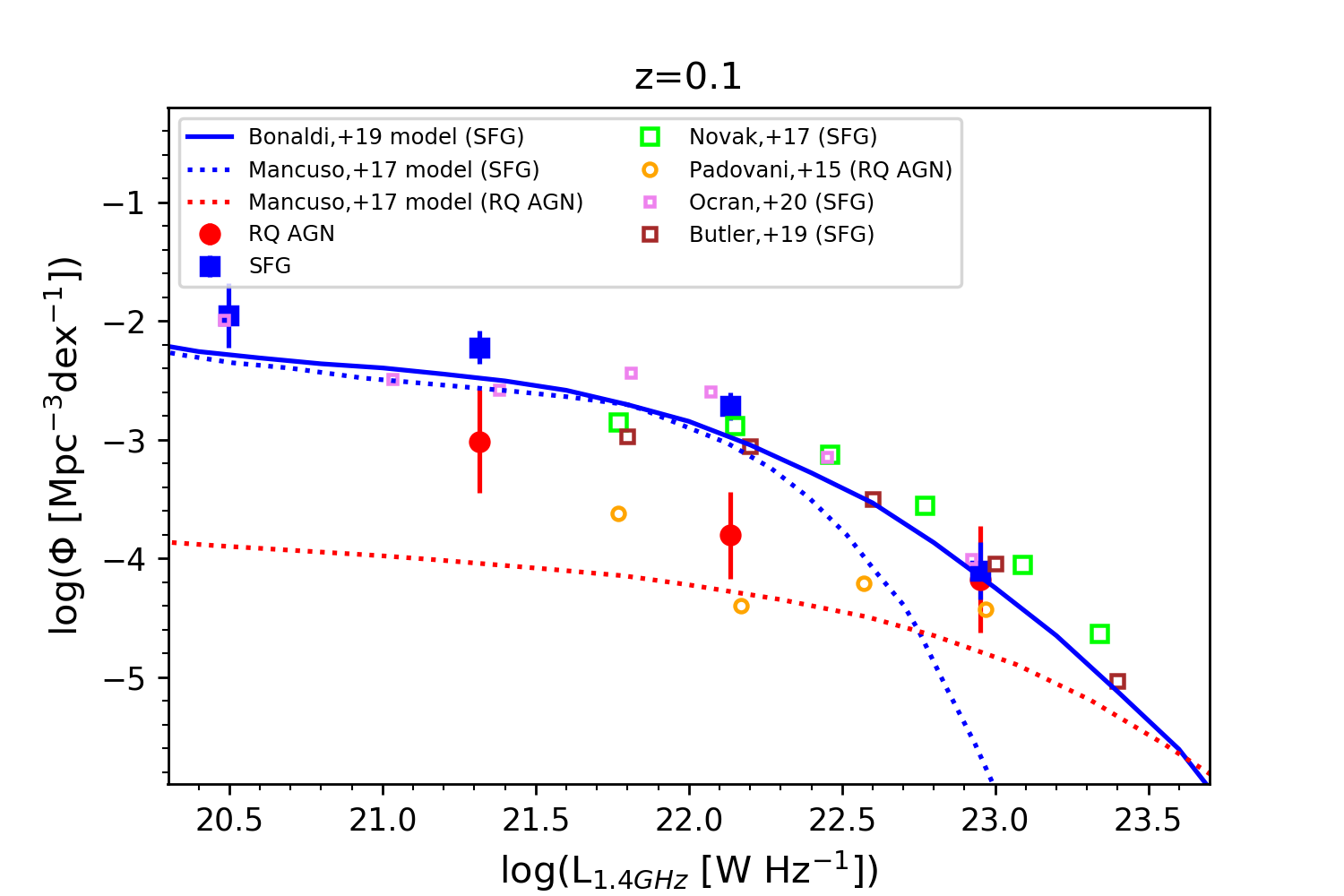}
\includegraphics[width=0.49\textwidth]{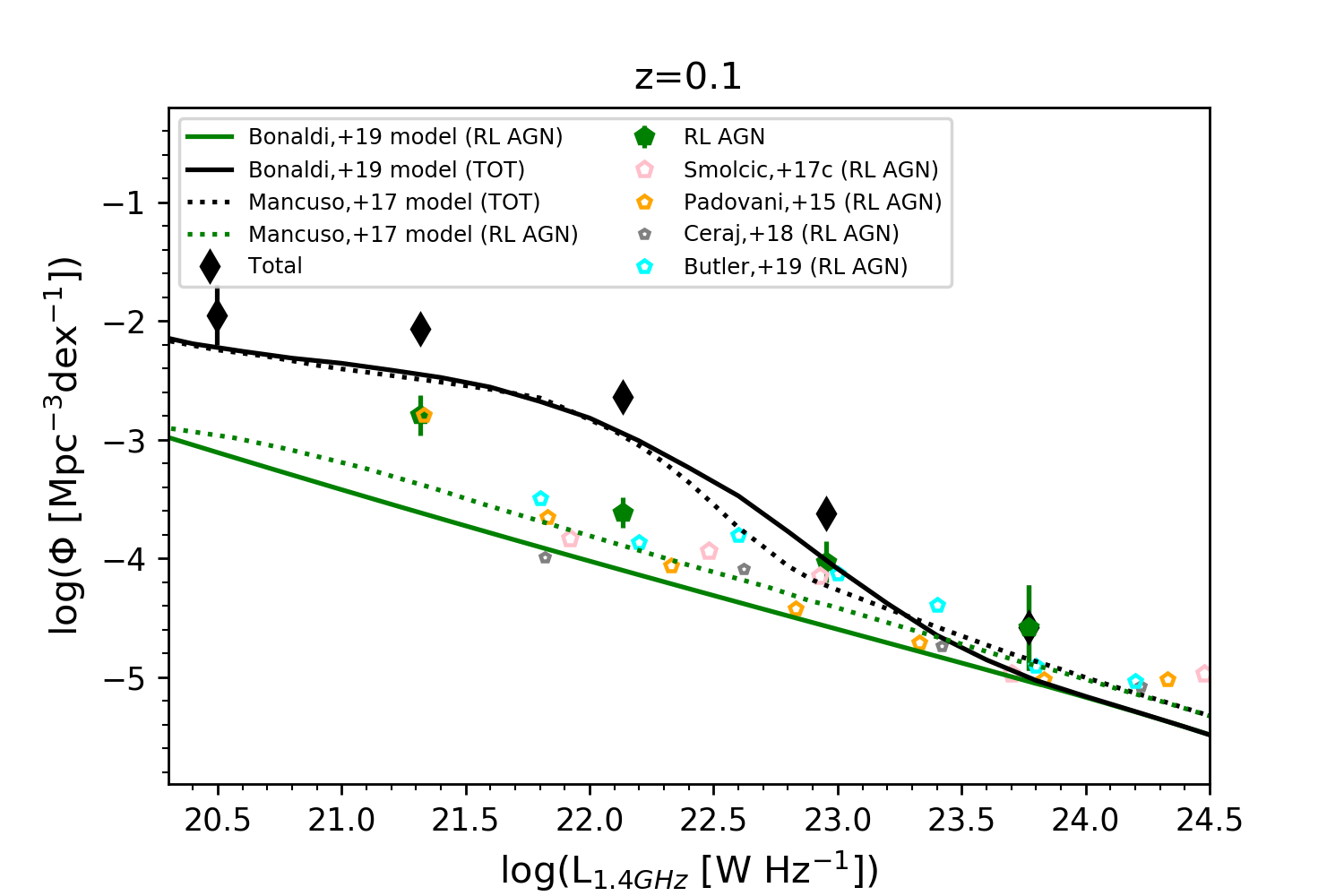}
\includegraphics[width=0.49\textwidth]{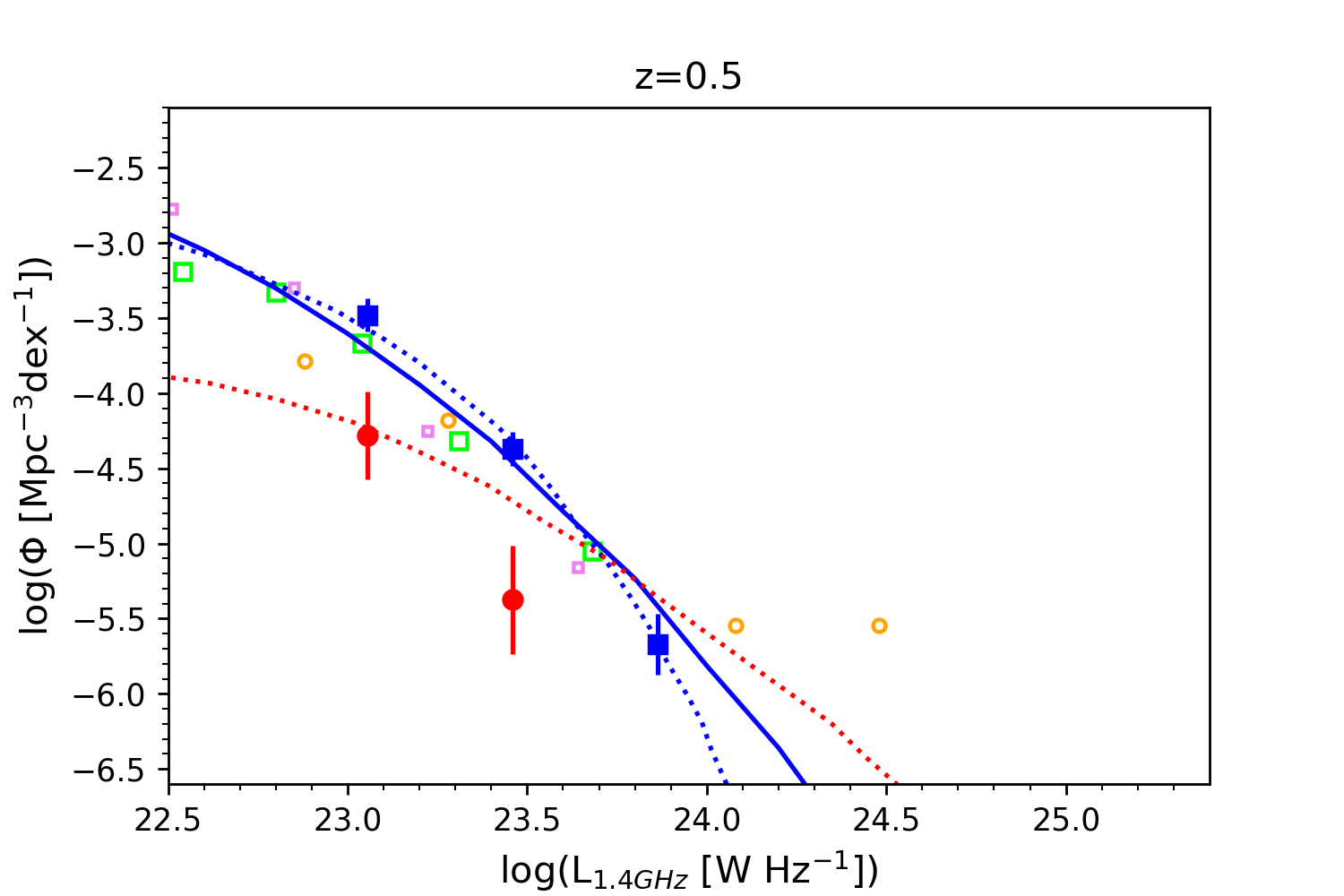}
\includegraphics[width=0.49\textwidth]{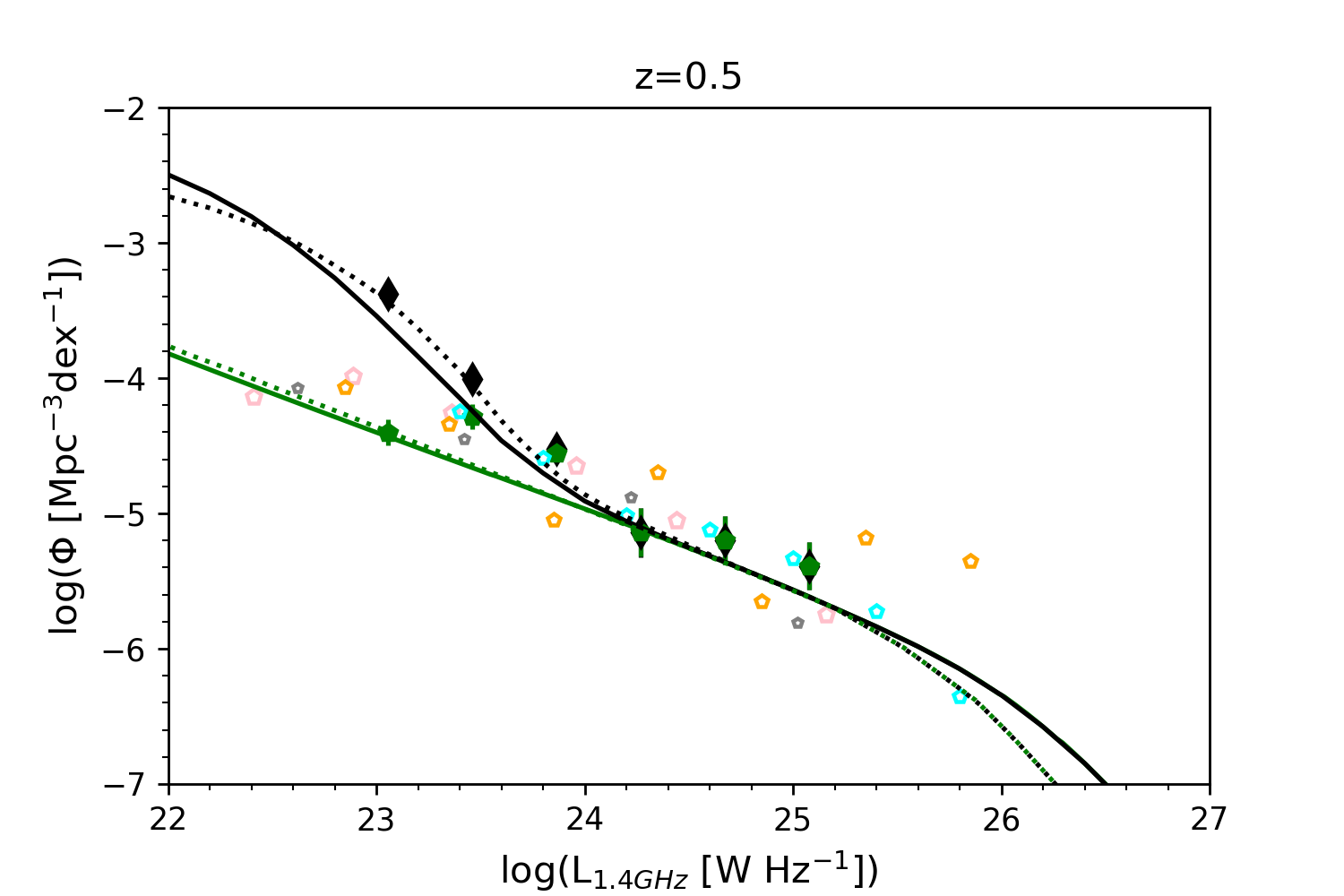}
\includegraphics[width=0.49\textwidth]{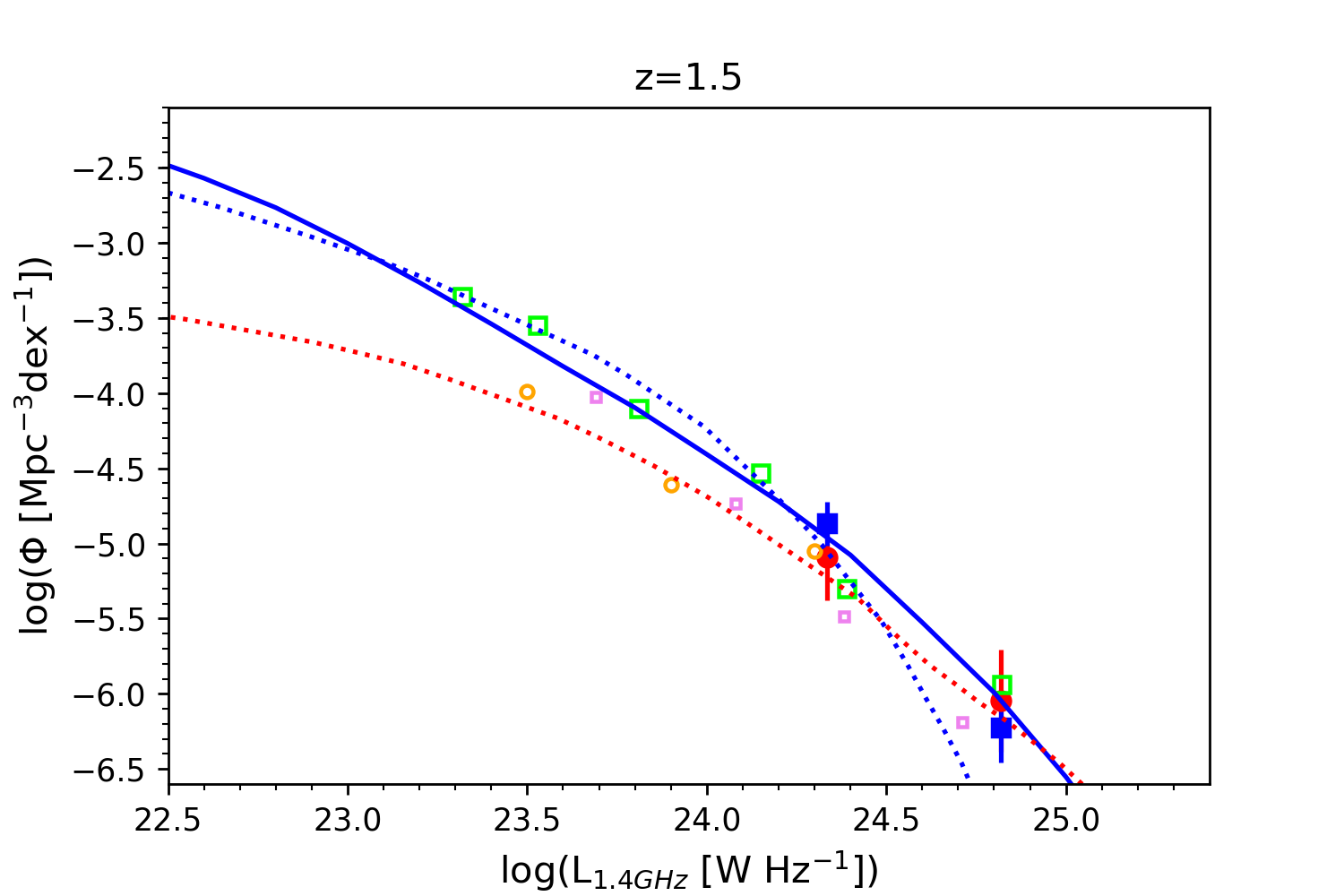}
\includegraphics[width=0.49\textwidth]{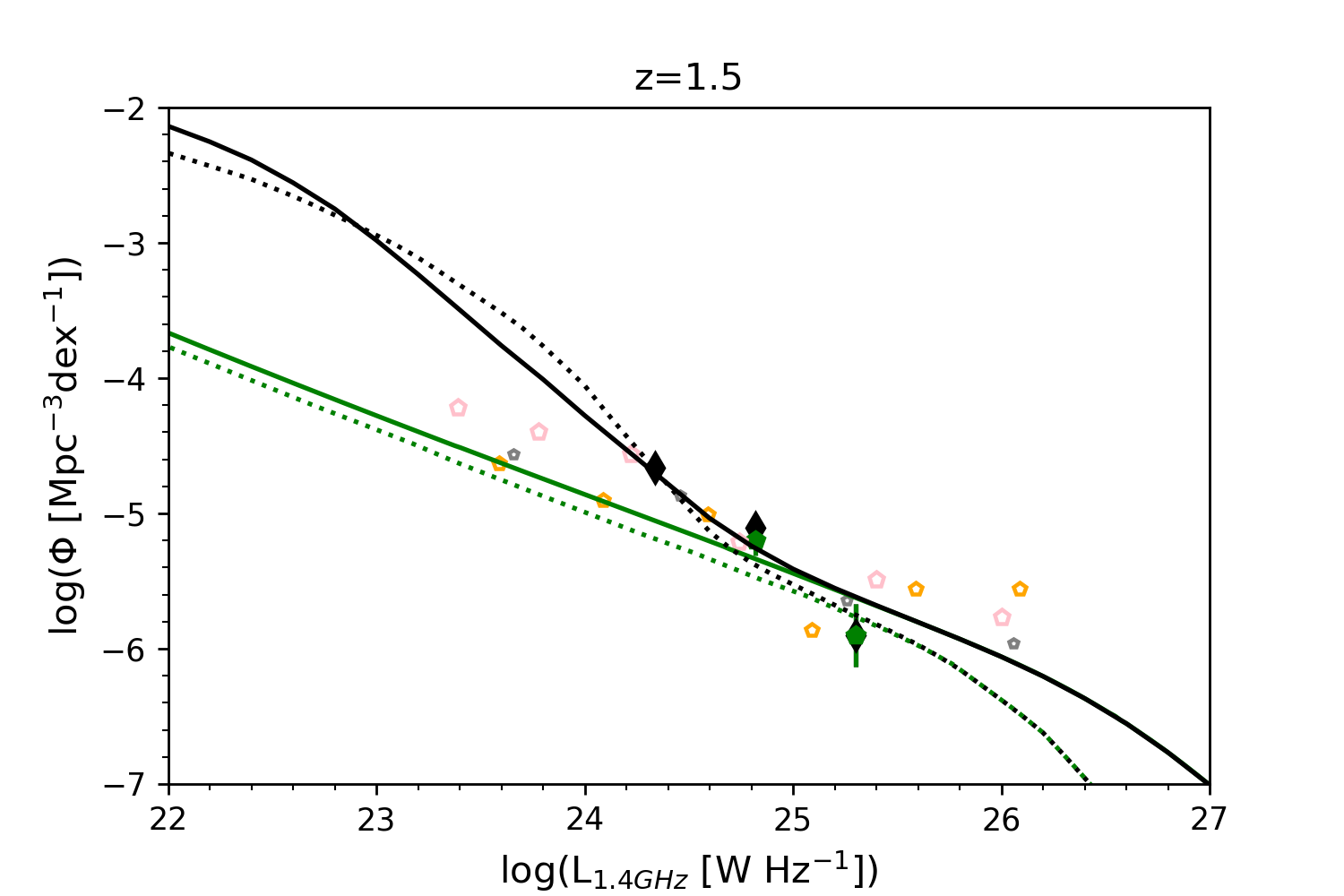}
\caption{Comparison of our estimates of the LFs of non-radio excess sources (SFGs and RQ AGNs)
and RL AGNs (left and right panels,
respectively) with those of SFGs by \citet{Novak17},
\citet[][including high-excitation sources whose radio emission likely originates from star formation]{Butler2019}
and \citet{Ocran2020},
of RQ AGNs by \citet{Padovani2015}
and of RL AGNs by \citet{Padovani2015}, \citet{Smolcic2017c}, \citet{Ceraj2018} and \citet{Butler2019}.
The estimates by \citet{Ocran2020} have been converted from 610\,MHz to 1.4\,GHz using the spectral index
($\alpha=0.8$) used by these authors.
Also shown, for comparison, are the models by \citet{Mancuso2017} for SFGs, RQ and RL AGNs
and by \citet{Bonaldi2019} for SFGs and RL AGNs.}
 \label{fig:LF_sub}
  \end{center}
\end{figure*}

The calculation of 1.4\,GHz luminosities from the measured flux densities
\begin{equation}\label{eq:Lnu}
L_\nu=\frac{4\pi d_L^2 S_\nu}{K(L_\nu,z)},
\end{equation}
$d_L$ being the luminosity distance, requires the evaluation of the K-correction
\begin{equation}
K(L_\nu,z)=\frac{(1+z)L[\nu(1+z)]}{L(\nu)}.
\end{equation}
We have adopted simple power-law spectra, $S_\nu \propto \nu^{-\alpha}$ so that
$K\propto (1+z)^{1-\alpha}$, with $\alpha=0.8$ for SFGs \citep{Condon1992,
Ibar2010} and RQ AGNs and $\alpha=0.75$ for RL AGNs
\citep[e.g.,][]{Smolcic2017}. The latter choice is obviously not appropriate
for flat-spectrum radio sources, which are however expected to be a small
minority.

The luminosity function (LF) in the $k$-th redshift bin was derived using the
$1/V_{\rm max}$ method \citep{Schmidt1968}:
\begin{equation}
\frac{dN(L_j,z_k)}{d\log L}=\frac{1}{\Delta \log L}\sum_{i=1}^{N_{j}} \frac{w_{i}}{V_{{\rm max},i}(z_k)}
\end{equation}
where $z_k$ is the bin center and the sum is over all the $N_{j}$ sources with
luminosity in the range [$\log L_{j}-\Delta\log L/2$, $\log L_{j}+\Delta\log
L/2$] within the redshift bin. We chose $\Delta z=0.2$ for $z < 2$ and $\Delta
z=0.4$ for $z > 2$. These redshift bins are large enough to have a sufficient
statistics and narrow enough to allow us to neglect evolutionary corrections
within the bin. The values of $\Delta\log L$ are different in different
redshift bins (see Table~\ref{tab:LF}); they are larger in the poorly populated luminosity ranges.

The weights $w_{i}$ take into account the incompleteness in the
identifications. It was evaluated from the ratio between the
$\log(S_{1.4\,GHz})$ distributions of all the sources and of the identified
sources, in each bin.
%
%
The $V_{{\rm max},i}$ is the comoving volume, within the solid angle of the
survey, enclosed between the lower limit, $z_{\rm min}$, of the bin and the
minimum between the upper limit, $z_{\rm max}$, and the maximum redshift at
which the source is above both the IRAC 2 ($44\,\mu$Jy) and the radio
($120\,\mu$Jy) flux limits.

The LFs of SFGs and of RQ AGNs were further corrected for misclassifications
based on the X-ray data, as described above. In the present data
situation the best we could do was to apply the correction factors of
sub-sect.~\ref{subsec:classification_SFG_RQAGN} to the normalizations at all
redshifts, including the corresponding uncertainty in the error budget.
Although the true correction factors are most likely redshift-dependent, they
are within our conservative error bars.

The Poisson error on $dN(L_j,z_k)/d\log L$ was estimated as:
\begin{equation}
\sigma_{j,k} = \left\{\sum_{i=1}^{N_{j}} \frac{w_{i}^{2}}{[V_{{\rm max},i}(z_k)]^{2}}\right\}^{1/2}.
\end{equation}
For the SFG and the RQ AGN populations the uncertainty is the sum in quadrature
of Poisson errors, of classification uncertainties and of the fluctuations of
the number of sources in each redshift and luminosity bin obtained from
simulations. For the RL AGNs and the total population, the uncertainty takes
into account the variance in the simulations and the Poisson errors.

Our estimates of the 1.4\,GHz LFs for each  source population are listed in
Table\,\ref{tab:LF}. Some examples are shown and compared with other estimates
and with models in Fig.\,\ref{fig:LF_sub}. RL AGNs have LFs much flatter than
SFGs, so that the two populations dominate at high and low radio luminosities,
respectively. The transition luminosity increases with $z$ from $L_{1.4\,\rm
GHz} \simeq 10^{23.2}$ to $\simeq 10^{24}\,\hbox{W}\,\hbox{Hz}^{-1}$.

At all redshifts the RQ AGNs have space densities comparable to SFGs for
$L_{1.4\,\rm GHz}> 10^{24}\,\hbox{W}\,\hbox{Hz}^{-1}$. The redshift
independence of the density ratios of the two populations implies similar
evolution, as previously noted by \citet{Padovani2015}. At lower radio
luminosities, however, SFGs outnumber RQ AGNs by a substantial factor. Part of
the difference may be due to the difficulty of pinpointing faint AGNs.

The consistency with earlier estimates and with  model predictions is generally
good.

\begin{table*}
\centering
\footnotesize
\caption{{Estimates of the 1.4\,GHz luminosity function, $\Phi=dN/(d\log L\, dV) [\hbox{Mpc}^{-3}]$, for RL AGNs, SFGs
and RQ AGNs, and total. On the right of the luminosity functions
we give the number, $N$, of sources in each bin. It includes the contributions of unidentified
and unclassified sources as well as the correction for the misclassification indicated
by X-ray data (for SFGs and RQ AGNs). The errors are dominated by classification uncertainties.}}
\begin{tabular}{ccccccccc}
\hline
\hline
$\log L$ & $\log(\Phi_{\rm TOT})$ & ${\rm N}_{\rm TOT}$ & $\log(\Phi_{\rm RL\, AGN})$ & ${\rm N}_{\rm RL\, AGN}$ & $\log(\Phi_{\rm SFG})$ & ${\rm N}_{\rm SFG}$ & $\log(\Phi_{\rm RQ\, AGN})$ & ${\rm N}_{\rm RQ\, AGN}$  \\
$[\hbox{W}\,\hbox{Hz}^{-1}]$ & $[\hbox{Mpc}^{-3}\,\hbox{dex}^{-1}]$ & & $[\hbox{Mpc}^{-3}\,\hbox{dex}^{-1}]$ & &
$[\hbox{Mpc}^{-3}\,\hbox{dex}^{-1}]$ & & $[\hbox{Mpc}^{-3}\,\hbox{dex}^{-1}]$ \\
\hline
\hline
\multicolumn{9}{c}{z = 0$-$ 0.2 } \\
20.50  &  -1.95 $\pm$ 0.25  &  4  &  $-$  &  $-$  &  -1.95 $\pm$ 0.27  &  4  &  $-$  &  $-$  \\
21.32  &  -2.06 $\pm$ 0.08  &  40  &  -2.79 $\pm$ 0.17  &  7  &  -2.22 $\pm$ 0.14  &  30  &  -3.01 $\pm$ 0.43  &  3  \\
22.13  &  -2.64 $\pm$ 0.05  &  96  &  -3.61 $\pm$ 0.13  &  11  &  -2.72 $\pm$ 0.12  &  78  &  -3.80 $\pm$ 0.36  &  7  \\
22.95  &  -3.62 $\pm$ 0.12  &  14  &  -4.03 $\pm$ 0.17  &  5  &  -4.10 $\pm$ 0.24  &  5  &  -4.18 $\pm$ 0.45  &  4  \\
23.77  &  -4.58 $\pm$ 0.36  &  2  &  -4.58 $\pm$ 0.36  &  2  &  $-$  &  $-$  &  $-$  &  $-$  \\
\hline
\multicolumn{9}{c}{z = 0.2 $-$ 0.4 } \\
22.61  &  -2.94 $\pm$ 0.11  &  96  &  -4.77 $\pm$ 0.17  &  6  &  -3.00 $\pm$ 0.16  &  78  &  -3.90 $\pm$ 0.38  &  12  \\
23.51  &  -4.26 $\pm$ 0.10  &  22  &  -4.52 $\pm$ 0.14  &  12  &  -4.79 $\pm$ 0.19  &  7  &  -5.08 $\pm$ 0.39  &  3  \\
24.41  &  -5.25 $\pm$ 0.28  &  2  &  -5.25 $\pm$ 0.28  &  2  &  $-$  &  $-$  &  $-$  &  $-$  \\
25.31  &  -5.62 $\pm$ 0.39  &  1  &  -5.62 $\pm$ 0.39  &  1  &  $-$  &  $-$  &  $-$  &  $-$  \\
\hline
\multicolumn{9}{c}{z = 0.4 $-$ 0.6 } \\
23.06  &  -3.37 $\pm$ 0.04  &  67  &  -4.40 $\pm$ 0.10  &  10  &  -3.48 $\pm$ 0.11  &  47  &  -4.28 $\pm$ 0.29  &  10  \\
23.46  &  -4.01 $\pm$ 0.05  &  31  &  -4.29 $\pm$ 0.09  &  13  &  -4.37 $\pm$ 0.11  &  17  &  -5.37 $\pm$ 0.36  &  1  \\
23.86  &  -4.52 $\pm$ 0.07  &  12  &  -4.56 $\pm$ 0.07  &  11  &  -5.67 $\pm$ 0.20  &  1  &  $-$  &  $-$  \\
24.27  &  -5.14 $\pm$ 0.18  &  2  &  -5.14 $\pm$ 0.18  &  2  &  $-$  &  $-$  &  $-$  &  $-$  \\
24.67  &  -5.20 $\pm$ 0.18  &  1  &  -5.20 $\pm$ 0.18  &  1  &  $-$  &  $-$  &  $-$  &  $-$  \\
25.08  &  -5.39 $\pm$ 0.18  &  2  &  -5.39 $\pm$ 0.18  &  2  &  $-$  &  $-$  &  $-$  &  $-$  \\
\hline
\multicolumn{9}{c}{z = 0.6 $-$ 0.8 } \\
23.49  &  -3.69 $\pm$ 0.04  &  77  &  -4.38 $\pm$ 0.08  &  18  &  -3.97 $\pm$ 0.12  &  39  &  -4.28 $\pm$ 0.28  &  20  \\
23.98  &  -4.37 $\pm$ 0.06  &  28  &  -4.41 $\pm$ 0.06  &  25  &  $-$  &  $-$  &  -5.45 $\pm$ 0.31  &  3  \\
24.47  &  -4.75 $\pm$ 0.07  &  13  &  -4.75 $\pm$ 0.08  &  13  &  $-$  &  $-$  &  $-$  &  $-$  \\
24.96  &  -5.01 $\pm$ 0.10  &  7  &  -5.01 $\pm$ 0.10  &  7  &  $-$  &  $-$  &  $-$  &  $-$  \\
25.94  &  -5.59 $\pm$ 0.15  &  2  &  -5.59 $\pm$ 0.15  &  2  &  $-$  &  $-$  &  $-$  &  $-$  \\
\hline
\multicolumn{9}{c}{z = 0.8 $-$ 1.0 } \\
24.00  &  -4.15 $\pm$ 0.08  &  81  &  -4.58 $\pm$ 0.09  &  34  &  -6.21 $\pm$ 0.38  &  1  &  -4.36 $\pm$ 0.29  &  46  \\
24.83  &  -5.17 $\pm$ 0.13  &  11  &  -5.22 $\pm$ 0.14  &  10  &  $-$  &  $-$  &  -6.12 $\pm$ 0.45  &  1  \\
25.67  &  -5.65 $\pm$ 0.21  &  4  &  -5.65 $\pm$ 0.21  &  4  &  $-$  &  $-$  &  $-$  &  $-$  \\
26.50  &  -6.24 $\pm$ 0.36  &  1  &  -6.24 $\pm$ 0.36  &  1  &  $-$  &  $-$  &  $-$  &  $-$  \\
\hline
\multicolumn{9}{c}{z = 1.0 $-$ 1.2 } \\
24.20  &  -4.49 $\pm$ 0.07  &  54  &  -4.83 $\pm$ 0.10  &  24  &  -4.94 $\pm$ 0.16  &  19  &  -5.22 $\pm$ 0.31  &  11  \\
25.02  &  -5.32 $\pm$ 0.13  &  9  &  -5.32 $\pm$ 0.13  &  9  &  $-$  &  $-$  &  $-$  &  $-$  \\
25.85  &  -5.60 $\pm$ 0.24  &  3  &  -5.60 $\pm$ 0.24  &  3  &  $-$  &  $-$  &  $-$  &  $-$  \\
\hline
\multicolumn{9}{c}{z = 1.2 $-$ 1.4 } \\
24.24  &  -4.41 $\pm$ 0.11  &  31  &  -5.33 $\pm$ 0.15  &  8  &  -4.53 $\pm$ 0.17  &  17  &  -5.33 $\pm$ 0.31  &  6  \\
24.86  &  -5.69 $\pm$ 0.21  &  2  &  -5.69 $\pm$ 0.21  &  2  &  $-$  &  $-$  &  $-$  &  $-$  \\
26.11  &  -6.24 $\pm$ 0.27  &  1  &  -6.24 $\pm$ 0.27  &  1  &  $-$  &  $-$  &  $-$  &  $-$  \\
\hline
\multicolumn{9}{c}{z = 1.4 $-$ 1.6 } \\
24.34  &  -4.66 $\pm$ 0.09  &  21  &  $-$  &  $-$  &  -4.87 $\pm$ 0.15  &  10  &  -5.09 $\pm$ 0.28  &  11  \\
24.82  &  -5.11 $\pm$ 0.10  &  8  &  -5.20 $\pm$ 0.11  &  6  &  -6.22 $\pm$ 0.23  &  1  &  -6.05 $\pm$ 0.34  &  1  \\
25.30  &  -5.90 $\pm$ 0.23  &  2  &  -5.90 $\pm$ 0.23  &  2  &  $-$  &  $-$  &  $-$  &  $-$  \\
\hline
\multicolumn{9}{c}{z = 1.6 $-$ 1.8 } \\
24.68  &  -4.87 $\pm$ 0.10  &  29  &  -5.42 $\pm$ 0.20  &  6  &  -5.19 $\pm$ 0.17  &  14  &  -5.48 $\pm$ 0.32  &  9  \\
25.55  &  -6.21 $\pm$ 0.39  &  1  &  -6.21 $\pm$ 0.39  &  1  &  $-$  &  $-$  &  $-$  &  $-$  \\
26.42  &  -6.14 $\pm$ 0.27  &  2  &  -6.14 $\pm$ 0.27  &  2  &  $-$  &  $-$  &  $-$  &  $-$  \\
\hline
\multicolumn{9}{c}{z = 1.8 $-$ 2.0 } \\
24.51  &  -4.76 $\pm$ 0.06  &  16  &  -5.18 $\pm$ 0.10  &  4  &  -5.19 $\pm$ 0.13  &  7  &  -5.38 $\pm$ 0.29  &  5  \\
24.85  &  -5.50 $\pm$ 0.09  &  4  &  $-$  &  $-$  &  -5.50 $\pm$ 0.13  &  4  &  $-$  &  $-$  \\
\hline
\multicolumn{9}{c}{z = 2.0 $-$ 2.4 } \\
24.78  &  -5.49 $\pm$ 0.12  &  12  &  -6.28 $\pm$ 0.33  &  2  &  -5.94 $\pm$ 0.19  &  4  &  -5.82 $\pm$ 0.30  &  6  \\
26.61  &  -6.62 $\pm$ 0.26  &  1  &  -6.62 $\pm$ 0.27  &  1  &  $-$  &  $-$  &  $-$  &  $-$  \\
\hline
\multicolumn{9}{c}{z = 2.4 $-$ 2.8 } \\
24.94  &  -5.51 $\pm$ 0.10  &  9  &  -6.09 $\pm$ 0.24  &  2  &  $-$  &  $-$  &  -5.64 $\pm$ 0.28  &  7  \\
26.23  &  -6.41 $\pm$ 0.19  &  1  &  -6.41 $\pm$ 0.19  &  1  &  $-$  &  $-$  &  $-$  &  $-$  \\
\hline
\multicolumn{9}{c}{z = 2.8 $-$ 3.2 } \\
25.02  &  -6.04 $\pm$ 0.21  &  3  &  $-$  &  $-$  &  $-$  &  $-$  &  -6.04 $\pm$ 0.30  &  3  \\
25.45  &  -6.02 $\pm$ 0.20  &  1  &  -6.02 $\pm$ 0.20  &  1  &  $-$  &  $-$  &  $-$  &  $-$  \\
\hline
\hline
\end{tabular}
\label{tab:LF}
\end{table*}

\begin{figure*}
\begin{center}
\includegraphics[width=0.49\textwidth]{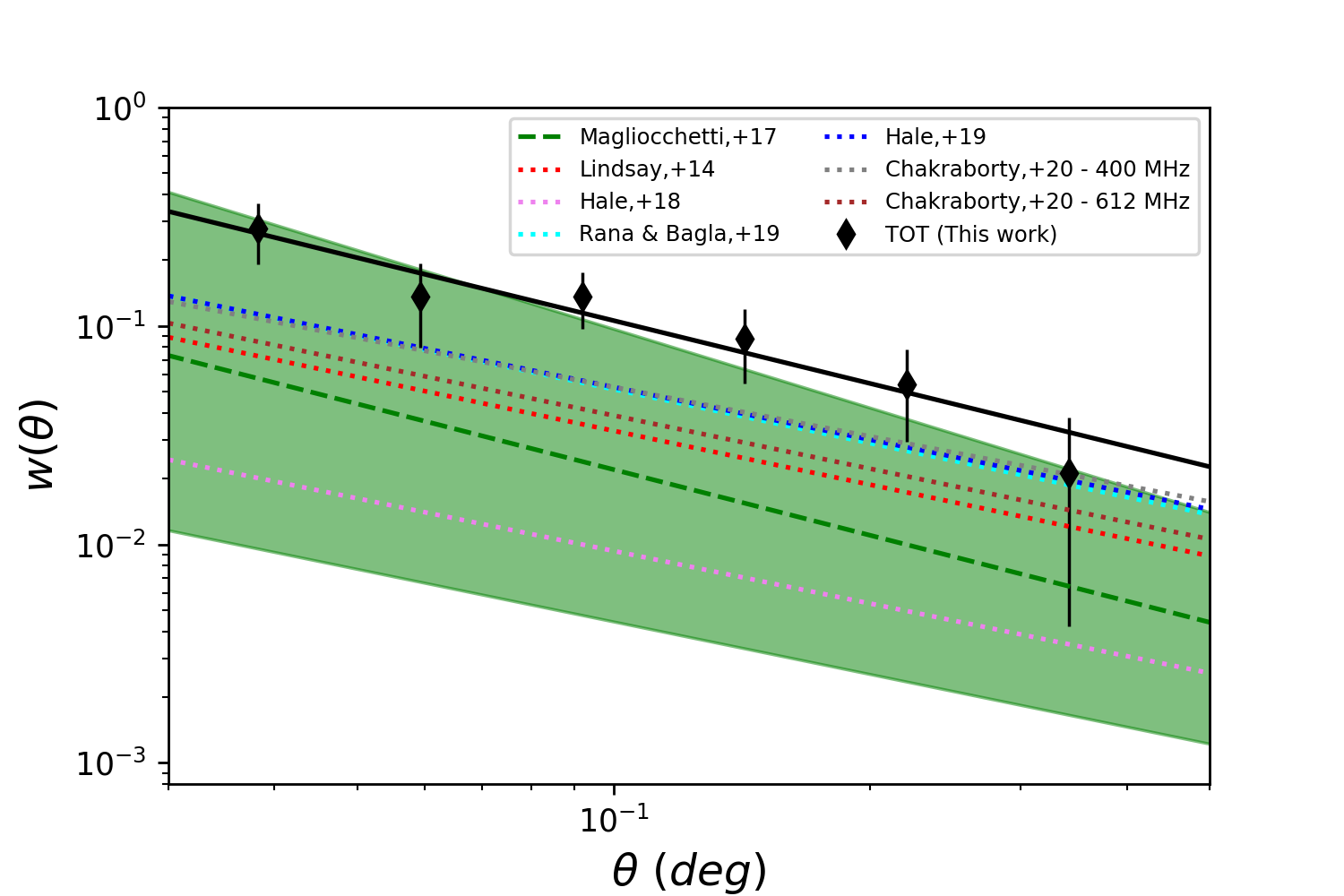}
\includegraphics[width=0.49\textwidth]{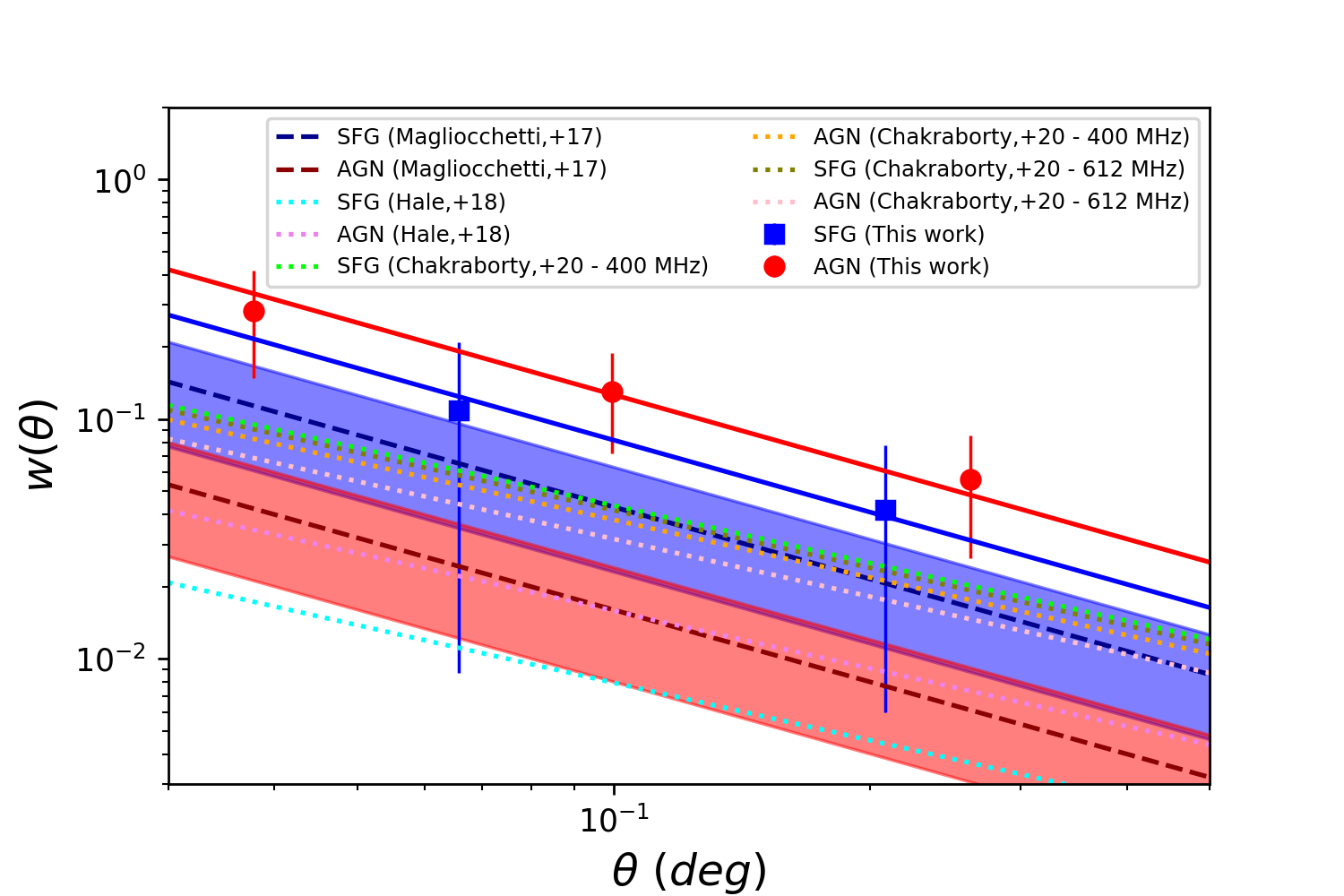}
\caption{Two-point angular correlation function, $w(\theta)$, derived from our
catalog (the datapoints), compared to the results of \citet{Magliocchetti2017} (the coloured bands). The left and
the right panels show, respectively, the $w(\theta)$ for the full $S_{1.4\,\rm GHz}$ sample
and for  ``SFGs'' and RL AGNs (``AGN'') separately. The poor statistics does not allow us
to derive separate correlation functions for SFGs and RQ AGNs.  Also \citet{Magliocchetti2017}
have considered the two populations together.
We show only bins where we have sufficient statistics and at angular scales $>$100\,arcsec (i.e.
larger than the maximum angular separation used to identify multi-component sources).
We have also added, for comparison, the best fit $w(\theta)$ reported by
other studies: \citet{Lindsay2014}, \citet{Hale2018}, \citet{Hale2019}, \citet{RanaBagla2019} and
\citet{Chakraborty2020}. The results for ``SFG'' and ``AGN'' by \citet{Hale2018} are the
fits given in their Table\,1.}
 \label{fig:clustering}
  \end{center}
\end{figure*}

\begin{figure*}
\begin{center}
\includegraphics[width=0.49\textwidth]{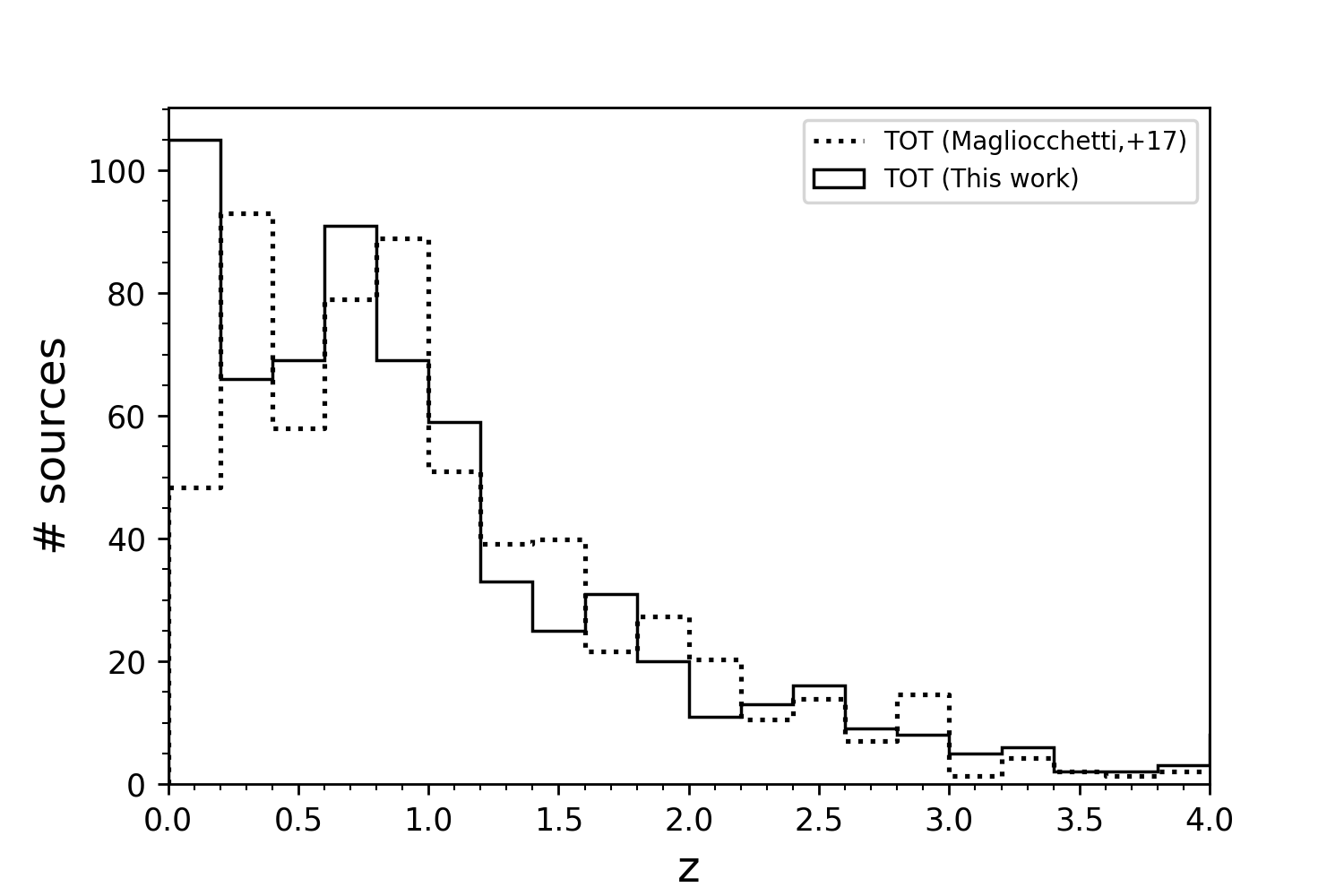}
\includegraphics[width=0.49\textwidth]{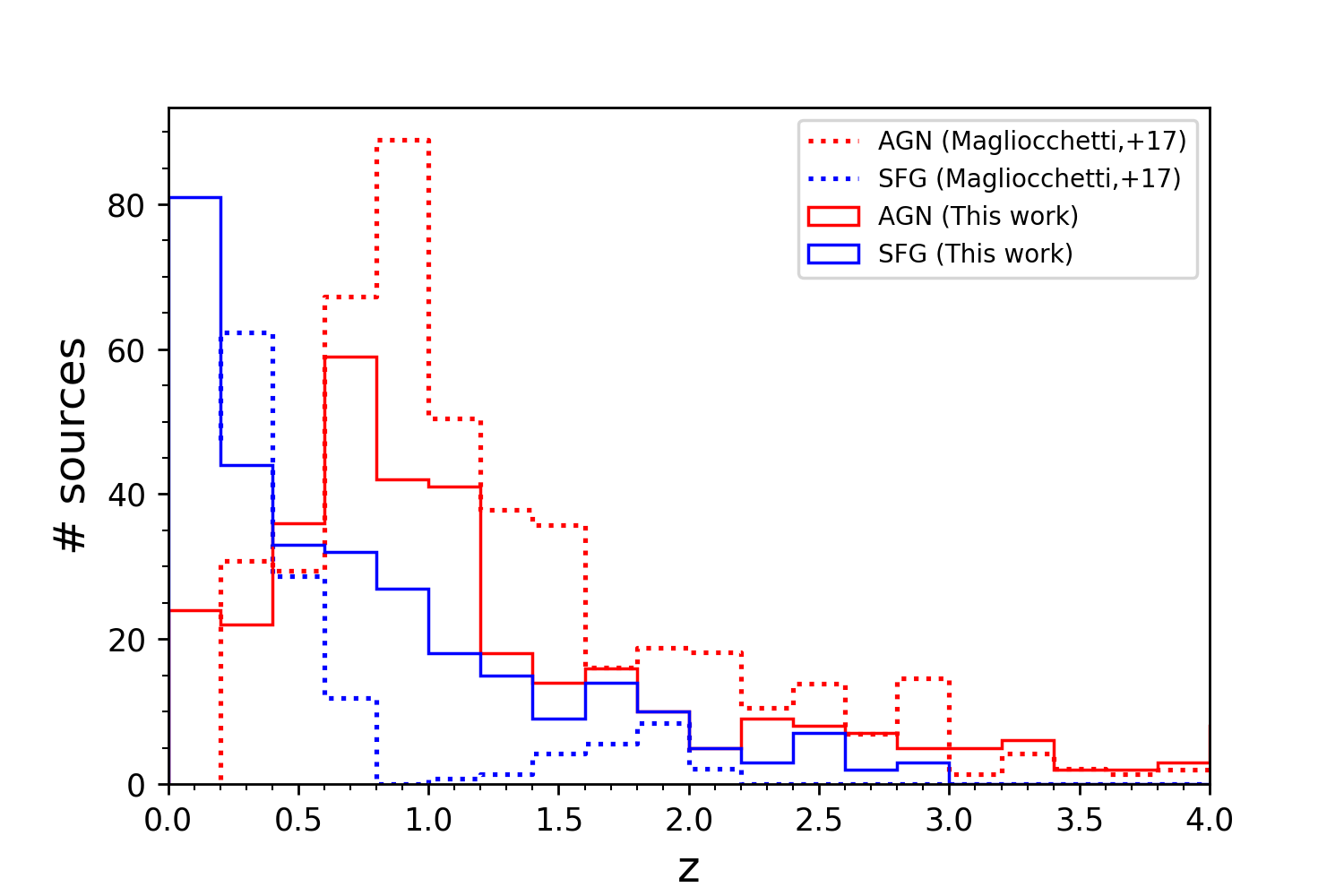}
\caption{Comparison between the redshift distributions of our sources brighter than
150\,$\mu$Jy and the \citet{Magliocchetti2017} one, normalized to the area of
our survey (i.e. $\sim$1.4 deg$^{2}$). The left panel shows the global
redshift distributions while the right one shows those of ``AGNs'' and of ``SFGs''
(see caption of Fig.~\protect\ref{fig:clustering}) separately.}
 \label{fig:z_distributions_M17}
  \end{center}
\end{figure*}

\subsection{Clustering}\label{subsec:clustering}

Measurements of the clustering properties of extragalactic source
populations are an efficient way to gain information on the relationship
between their distribution and that of dark matter matter halos, hence on their
formation history and on their environment \citep{Peebles1980}. The radio band
has important advantages over the optical band where most clustering studies
have been carried out. It is better suited to investigate the large scale
structure at high redshift thanks to the strong cosmological evolution of radio
loud AGNs and to the immunity to dust obscuration. Also, the survey speed is
generally higher in the radio making much easier to uniformly cover large areas
of the sky. This has motivated many clustering studies at various radio
frequencies and to different depths \citep[e.g.][]{Lindsay2014,
Magliocchetti2017, Hale2018, Hale2019, RanaBagla2019, Chakraborty2020}.

We have computed the 2-point angular correlation function, $w(\theta)$,
using the public Python package AstroML \citep{astroML, astroMLText} and
specifically the ``bootstrap\_two\_point\_angular''
function\footnote{\url{https://www.astroml.org/modules/generated/astroML.correlation.bootstrap_two_point_angular.html}}.
The $w(\theta)$ is computed exploiting the  Landy-Szalay estimator
(\citealt{Landy1993}):
\begin{equation}
w(\theta)=\frac{DD(\theta)-2DR(\theta) + RR(\theta)}{RR(\theta)},
\end{equation}
where $DD(\theta)$ is the real number of source pairs separated by an angle
$\theta$, $RR(\theta)$ is the number of pairs expected in the case of a random
distribution and $DR(\theta)$ is the number of data-random pairs. We limited
ourselves to angular scales $> 100\,$arcsec since smaller scales may be
affected by an over/under-correction for multiple source components. The
results are presented in Fig.\,\ref{fig:clustering} where they are also
compared with previous estimates.

The most straightforward comparison is that with the estimate by
\citet{Magliocchetti2017} who used a sample of similar size (they have 968
sources) and depth to ours, selected at the same frequency. To put ourselves on
an exactly equal footing we have adopted their 1.4\,GHz flux density threshold
(0.15\,mJy).


On the scales probed by our sample, the $w(\theta)$ is well described by a power
law, $w(\theta)=A (\theta/\hbox{deg})^{1-\gamma}$ with $A\simeq
(0.012\pm0.005)$ and $\gamma = 1.957 \pm 0.151$. Our estimate is higher than
that by \citet{Magliocchetti2017}, although consistent with it within the
errors. Part of the difference may be due to sample variance.

Our global redshift distribution of sources with $S_{1.4\,\rm GHz}\ge
150\,\mu$Jy is similar to that by
\citet[][Fig.\,\ref{fig:z_distributions_M17}]{Magliocchetti2017}  who have a
higher fraction of spectroscopic plus reliable photometric redshifts ($\simeq
92\%$). This lends support to the validity of our statistical procedure for
redshift assignments.

The determination of the correlation function of each source population is
hampered by the poor statistics which prevents a simultaneous estimate of both
$A$ and $\gamma$. Fixing $\gamma$ to the best fit value for the full sample
($\gamma=1.957$), we get the results shown in the right-hand panel of
Fig.\,\ref{fig:clustering}. This figure shows that the difference with
\citet{Magliocchetti2017} mostly arises from  RL AGNs, for which our estimate
is significantly higher.

The origin of this difference can be traced back to the criteria used to
separate ``AGNs'' from ``SFGs''. The separation made by
\citet{Magliocchetti2017} relied purely on radio luminosity, independently of
the SFR. The result is that, except for a small secondary peak at $z\simeq
1.9$, their ``SFGs'' are at $z\le 0.8$ while in our case a substantial fraction
of them is located at $0.6 \le z \le 1.5$ with a tail extending up to $z\simeq
3$, mostly populated by RQ AGNs.

According to our estimate, RL AGNs and SFGs have correlation functions with
amplitudes  $A=0.013\pm0.001$ and $0.008\pm0.001$, respectively. In contrast,
\citet{Magliocchetti2017} found a somewhat stronger clustering for ``SFGs''
than for ``AGNs''.

The other determinations of the clustering amplitude shown in
Fig.\,\ref{fig:clustering} are consistent within the errors, with our results,
although somewhat lower. The comparison however has to be taken with caution
since they refer to surveys at different frequencies and/or of different
depths. In fact, the mixture of source populations (e.g. the relative fractions
of steep- and flat-spectrum sources) vary with frequency and the redshift
distributions vary with the survey depth. Indeed, variation of the clustering
amplitude with flux density limit has been reported by \citet{RanaBagla2019}.

\citet{Lindsay2014} studied the clustering properties of sources in the
Faint Images of the Radio Sky at Twenty-cm (FIRST) survey at 1.4\,GHz, with a
flux limit of 1\,mJy, several times higher than that of our sample.
\citet{Hale2018} investigated a sample selected at 3\,GHz flux limited at
$\simeq 13\,\mu$Jy. \citet{Hale2019} analyzed a sample of sources brighter than
$\sim 2\,$mJy at 144\,MHz. \citet{RanaBagla2019} used a shallow large-area
sample selected at 150\,MHz and used various flux density thresholds (50, 60,
100 and 200\,mJy). \citet{Chakraborty2020} used 400\,MHz observations flux
limited at $100\,\mu$Jy and 612\,MHz data down to $50\,\mu$Jy.



\section{Conclusions}\label{sec:conclusions}

We have exploited the rich multi-band information available on the LH field to
classify radio sources with $S_{1.4\,\rm GHz}> 120\,\mu$Jy detected by the deep
1.4\,GHz WSRT survey (rms noise of $11\,\mu$Jy/beam) over an area of $\simeq
1.4\,\hbox{deg}^{2}$. After correcting for multi-component sources, our sample
comprises 1173 objects. Deep multi-band observations are available for 921
(IRAC identified) sources.

Spectroscopic or reliable photometric redshifts are available for 744 of our
sources. Using these data we found tight anti-correlations between IRAC band\,1
($3.6\,\mu$m) or band\,2 ($4.5\,\mu$m) and $\log z$. These were exploited to
assign redshifts to the remaining 177 IRAC identified sources. The spread
around the mean relations was taken into account using a Monte Carlo approach.

Radio-excess sources, i.e. sources with $\log(L_{\rm 1.4GHz}/\hbox{SFR})$ above
the redshift-dependent threshold defined by \citet{Smolcic2017b}, were
classified as RL AGNs. The few objects for which no estimate of the SFR was
possible were classified on the basis of their radio luminosity, following
\citet{Magliocchetti2017}. In total, RL AGNs comprise 471 sources ($\simeq
40\%$ of the sample).

The other sources were classified as either SFGs or RQ AGNs using the
redshift-dependent criterion or the $K_s$--IRAC diagnostic plot by
\citet{Messias2012} or the IRAC colour-colour diagram \citep{Donley2012},
depending on the available data. We got 310 SFGs and 124 RQ AGNs. Only 16
IRAC identified sources couldn't be classified.

Our classification was checked using the deep X-ray survey with
XMM--\textit{Newton} \citep{Brunner2008} of $\simeq 10\%$ of the field. Three
out of the 24 sources ($\sim$13\%) in this area classified as SFGs have X-ray
luminosity in excess of expectations from star formation, suggesting that they
are misclassified RQ AGNs. We have therefore decreased by $\sim$13\% the total number
of SFGs and correspondingly increased that of RQ AGNs.

We have computed the differential number counts and the luminosity functions at
several redshifts of each population and compared them with models and with
earlier estimates made using data from different surveys and applying different
approaches. A reassuring consistency with both models and other estimates was
found, although we find differences in some details.

Our results confirm that RL AGNs dominate the 1.4\,GHz counts above
$\sim300\,\mu$Jy, while SFGs$+$RQ AGNs take over at fainter flux densities. 

The deep counts of RQ AGNs and of SFGs have a similar shape, suggesting a
similar evolution, consistent with the results by \citet{Padovani2015}. At the
brightest radio luminosities, SFGs and RQ AGNs have similar space densities at
all redshifts, confirming that the evolution is similar. The ratio of RQ AGN to
SFG space densities drops rapidly with decreasing radio luminosity. This may
be, at least in part, the effect of the difficulty of identifying faint AGNs.

The amplitude of the angular correlation function of our sources is somewhat
larger than, but consistent within the errors with that found by
\citet{Magliocchetti2017} for a sample of the same depth. On the other hand,
the clustering of our RL AGNs is significantly stronger than theirs. We argue
that the difference can arise from the different classification criteria.
Our results are broadly consistent with clustering properties of radio
sources reported by previous studies. We caution however that the comparison
requires carefulness because of the different selection frequencies and/or survey
depths.

\section*{Acknowledgements}
Thanks are due to the anonymous referee for thoughtful comments that
helped improving the paper. We acknowledge support from INAF under PRIN
SKA/CTA FORECaST, from the Ministero degli Affari Esteri e della Cooperazione
Internazionale - Direzione Generale per la Promozione del Sistema Paese
Progetto di Grande Rilevanza ZA18GR02 and the South African Department of
Science and Technology's National Research Foundation (DST-NRF Grant Number
113121). Marisa Brienza acknowledges support from the ERC-Stg DRANOEL, no
714245.

\section*{Data availability}
The data underlying this article are available in the article and in its online supplementary material.

\bibliographystyle{mnras}
\bibliography{biblio} 

\bsp	
\label{lastpage}
\end{document}